\documentclass[12pt,letterpaper]{article}
 
\usepackage{graphicx}
\usepackage{subfig}

\usepackage{ifpdf}

\usepackage{rangecite}

\usepackage{amssymb}
\usepackage{amsmath}
\allowdisplaybreaks[1]

\usepackage{longtable}

\def\be{\begin{equation}}
\def\ee{\end{equation}}
\def\bear{\begin{eqnarray}}
\def\eear{\end{eqnarray}}
\def\nn{\nonumber}

\def\half{{{1\over 2}}}

\newcommand{\ket}[1]{\left|#1\right>}
\newcommand{\bra}[1]{\left<#1\right|}
\newcommand{\al}{\alpha_{0}}

\newcommand{\Nt}{\tilde{N}}

\mathchardef\mhyphen="2D

\begin{document}

%%%%%%%%%%%%%%%%%%%%%%%%%%%%%%%%%%%%%%%%%%%%%%%%%%%%%%%%%%%%%%%%%%%%
%  TITLE PAGE                                                      %
%%%%%%%%%%%%%%%%%%%%%%%%%%%%%%%%%%%%%%%%%%%%%%%%%%%%%%%%%%%%%%%%%%%%
%  Find all ***                                                    %
%%%%%%%%%%%%%%%%%%%%%%%%%%%%%%%%%%%%%%%%%%%%%%%%%%%%%%%%%%%%%%%%%%%%

\begin{titlepage}
~
\vskip 1in
\begin{center}
{\Large
{SFT on separated D-branes and D-brane translation}}
\vskip 0.5in
{Joanna L. Karczmarek and Matheson Longton}
\vskip 0.3in
{\it 
Department of Physics and Astronomy\\
University of British Columbia
Vancouver, Canada}
\end{center}

\vskip 0.5in
\begin{abstract}
We discuss novel properties of the string field and the Open String
Field Theory action arising in a system with multiple D-branes, then
use the level truncation scheme to study marginal deformations and 
tachyon condensation in a system with two parallel but separated 
branes.  We find string fields corresponding to D-brane decay
combined with a finite change in the distance between the 
two D-branes.  Using D-brane separation as a yardstick,
we are able to continuously control the spacetime displacement of
the D-branes and find that our solutions exist only for a finite 
range of this displacement. Thus, at least in level truncation, Open String 
Field Theory seems unable to describe the entire CFT moduli space.

\end{abstract}
\end{titlepage}

\tableofcontents

%%%%%%%%%%%%%%%%%%%%%%%%%%%%%%%%%%%%%%%%%%%%%%%%%%%%%%%%%%%%%%%%%%%%
%  DRAFTMORE ONLY; ADD MORE AUTHORS IF NEEDED                      %
%%%%%%%%%%%%%%%%%%%%%%%%%%%%%%%%%%%%%%%%%%%%%%%%%%%%%%%%%%%%%%%%%%%%

%%%%%%%%%%%%%%%%%%%%%%%%%%%%%%%%%%%%%%%%%%%%%%%%%%%%%%%%%%%%%%%%%%%%
%  BEGIN HERE                                                      %
%%%%%%%%%%%%%%%%%%%%%%%%%%%%%%%%%%%%%%%%%%%%%%%%%%%%%%%%%%%%%%%%%%%%

\section{Introduction}

String Field Theory has seen great progress in the last decade.
Different classes of solutions in cubic SFT---tachyon condensation, marginal
deformations and time dependent brane decay---have been understood
through either numerical studies in level truncation
\cite{Sen:1999nx,Gaiotto:2002wy,Sen:2000hx,Sen:2004cq,Moeller:2000jy}
or by constructing exact analytic solutions
\cite{Schnabl:2005gv,Ellwood:2006ba,Erler:2009uj,Schnabl:2007az,Kiermaier:2007ba,Kiermaier:2007vu,Kiermaier:2010cf},
or both (for a review, see for example 
\cite{Fuchs:2008cc}). 
With a few exceptions (see for example \cite{Ellwood:2006ba,Kluson:2002hr})
most of this work has focused on Open String Field Theory in the presence of
a single D-brane.  In this paper we begin an exploration of cubic OSFT in
the presence of multiple D-branes.  We find that the new degrees of freedom
corresponding to off-diagonal components of the Chan-Paton matrices
lead to new types of solutions.  These solutions provide us with new tools 
to explore the structure and properties of cubic OSFT.  

We work in bosonic string theory in 26 dimensions and focus on cubic OSFT in
the presence of two parallel D24-branes extended in $X_0$,$\ldots$,$X_{24}$ directions
and separated in the $X_{25}$ direction, studying solutions 
to this SFT as a function of the separation of the two D24-branes.  
The presence of two D-branes, as opposed to just one, implies that
each spacetime field of the string field is replaced by a 2$\times$2 matrix 
of fields\footnote{This is in some ways similar to the situation studied in
\cite{Bagchi:2008et}, which focused on a separated brane-antibrane
system.  One key difference is the unbroken SU(2) symmetry
present in our scenario at zero separation.}.  This matrix is either 
hermitian or antihermitian (see Section 
\ref{hermicity} for more details).  The diagonal elements of this matrix, 
which we will call here the 11 and the 22 elements, correspond
to strings that have both ends attached to either one of the two 
D-branes.  In addition, the string field contains off-diagonal fields,
12 and 21, corresponding to strings stretched 
between the two D-branes.  When the separation is zero, our SFT has a 
SU(2) symmetry which is apparent in the action, while at non-zero 
separation this symmetry is broken and the distinction between diagonal
and off-diagonal fields becomes important.  For the 11 and the 22 
fields, the action contains a copy of the single-brane 
OSFT action each.  
This, combined with the fact that the action has no terms linear in the
off-diagonal fields 12 and 21, implies that all the already-known solutions in 
single-brane OSFT exist for the two-brane configuration at any D-brane separation 
(with all the 12 and the 21 fields simply set to zero).  The cubic part of the action couples 
off-diagonal elements to the diagonal ones.  Thus, when the off-diagonal 
elements 12 and 21 are nonzero there exist new solutions which have 
no analog in the single-brane theory.  It is these solutions we
set out to study, using a level truncation scheme.  We find that these
solutions present an interesting interplay between tachyon
condensation and marginal deformations.

Since we have two D-branes, we can allow one to decay while the other does not.
For zero separation, there is a full SU(2) family of such
solutions \cite{Ellwood:2006ba}.  For non-zero separation, with a broken SU(2)
symmetry, we would expect only those solutions 
where either the 11 or the 22 tachyon field develops a vev to exist,
corresponding to the decay of either the first or the second D-brane.
However, the 11 and the 22 sectors of the string field each contain a massless string
state whose vev corresponds to a marginal deformation of the worldsheet
Conformal Field Theory that can be interpreted as displacing the D-branes 
from their original position in the $X_{25}$ direction
(T-dual to the Wilson line in that direction).  At any D-brane separation,
there should exist a string field with non-zero massless string field vevs
that physically corresponds to the two D-branes coming together to zero separation.
Once in this new configuration, should one of the D-branes decay, the decay would be happening
at the SU(2)-symmetric point.  Thus we would expect a full SU(2) family of solutions
that represent a combination of such D-brane translation with tachyon condensation.  
However, as it is not possible to fully restore the SU(2) symmetry in a 
level truncated model, we obtain instead isolated solutions
in which the off-diagonal 12 and 21 tachyon elements develop a non-zero 
vev together with the 11 and the 22 massless string elements.  

Thus, we find a family of solutions (one solution at every D-brane separation)
in which the off-diagonal tachyon field is non-zero.    
At zero D-brane separation, the vevs of the massless modes are zero, and our
solution corresponds to a SU(2) rotation of a diagonal solution in which one
of the two D-branes has decayed.  At non-zero separation, the massless string state 
vevs are non-zero and increase with increased D-brane separation.
Thus, we can interpret our solutions as a combination of
a marginal deformation (D-brane translation)
and D-brane decay.  These solutions have an energy comparable to that of a single D-brane
decay.

Our construction should be compared to that in \cite{Sen:2000hx,Sen:2004cq}.
There, once a finite vev for the field corresponding to
the massless string state was fixed, no solutions were 
found in level truncation; one equation of motion was left unsatisfied. 
We are able to find such solutions because in 
the untruncated theory there exists a continuously parametrized family of 
solutions, with an isolated solution point surviving the truncation.
By changing the original separation of the D-branes, 
we are able to find solutions to the equations of motion 
in which the deformation parameter has a finite vev which can be continuously adjusted.
The price we pay for this is the presence of tachyon condensation
which makes the marginal part of the deformation harder to isolate and analyze
(see Section \ref{decay:discussion} for details).

The presence of two D-branes in the picture gives us an important
tool: the physical separation of the D-branes is 
known and adjustable, thus it can be used as a yardstick for 
measuring the physical effect of the SFT deformation.
Previous work on computing the CFT vev corresponding to a given
SFT vev includes \cite{Sen:2004cq}, which compared the 
energy-momentum tensors in the CFT and the SFT approaches, and
\cite{Coletti:2003ai}, where lowest order corrections to the 
field redefinition between SFT fields and worldsheet fields
were found.  The solutions we find allow us to directly measure the vev of the conformal
field theory parameter (i.e., the physical position of the D-brane)
as a function of the deformation parameter in SFT.
The presence of tachyon condensation means that the
deformation parameter in our setup is not directly related to 
the vev of the SFT marginal parameter used in 
\cite{Sen:2000hx}, but the problem of connecting them has been
reduced to one purely within the SFT of one D-brane and
is left for future work.  A different approach to this problem is
described in Section \ref{su2}.

We show that there exists a maximum
separation between the D-branes, of order $\sqrt {\alpha '}$, beyond which no deformation
in OSFT is able to bring the two D-branes back together,
suggesting that it is not possible to cover the entire moduli space
of CFTs with OSFT in a single coordinate system.
We see a mechanism similar to that in \cite{Sen:2000hx,Sen:2004cq}, where
two branches of OSFT solutions merge with no real solutions in existence
beyond the point at which they meet.  Note that because our 
adjustable parameter is the distance between the D-branes and not
the infinitesimal marginal parameter for brane translation, our
conclusions are not affected by the possibility that the marginal parameter
might not parametrize the marginal trajectory in string field space 
beyond some finite distance.

This paper is organized as follows:  In Section \ref{SF}, we describe the 
properties of the string field in a scenario with $N$ parallel D24-branes.
We discuss in detail the reality condition, twist and Siegel gauge.
We find that both the reality condition and the twist even condition lead 
to novel consequences once multiple D-branes are present.
For example, the twist even condition is not equivalent to even level.
In Section \ref{truncation} we discuss level
truncation, in Section \ref{action} we construct the OSFT action
and in Section \ref{2branes} we discuss properties of the action peculiar to
the set-up with just 2 D-branes.  In Section 3, we present our solutions
corresponding to a combination of a marginal deformation and D-brane decay
and discuss their interpretation. 
Finally, in Section 4, we briefly discuss an attempt to restore SU(2) 
symmetry using a purely marginal deformation to bring separated D-branes back 
to coincidence.  We use $\alpha' = 1$ convention throughout the paper,
except where we restore $\alpha'$ explicitly for clarity.

\section{OSFT action for multiple D-branes}

In this section, we study the string field and the OSFT action in a
scenario with multiple D24-branes allowing for nonzero separation 
and for zero-mode fluctuations in the transverse direction.  
We find that the construction of a real, twist even, gauge 
fixed string field is
more involved than in the case of a single D-brane.  The complications
are caused by several new elements: the string field has more than one
vacuum (since there is one in every combination of Chan-Patton factors),
twist acts nontrivially on these multiple vacua, and the zero mode in
the transverse spacial direction takes different values depending 
the vacuum state. Our analysis would be equally applicable 
to other SFT scenarios where some (or all) of these elements are present.

\subsection{The real string field}
\label{SF}

Consider a stack of $N$ D24-branes, with $X^{25}$ as the transverse
direction.
The mode expansion for the $25^{th}$ coordinate of a string
starting on brane $i$ and ending on brane $j$ is
\be
X^{25} = y_i - {1\over 2\pi i}  (y_j - y_i) \ln (z/\bar z) +
i\sqrt{1 \over 2} \sum_{m\neq 0} {\alpha_m^{25} \over m}
\left ( {1\over z^m} - {1\over \bar z^m}\right )~,
\label{dirichlet}
\ee
where the $25^{th}$ dimension is non-compact and $y_i$ are the positions
of D24-branes in the $25^{th}$ dimension.
$\alpha^{25}_0$ can be nonzero when acting on a string field living 
on separated D-branes, when acting in a particular $ij$ sector,
we have:
\be
\alpha^{25}_0 ~\rightarrow~ d_{ij} ~\mathop{=}^{\mathrm{def}}~ - {y_j - y_i \over \pi \sqrt 2} ~.
\label{distance}
\ee

We assume standard mode expansions in the
other $25$ directions and in the ghost sector.  Since
the $25^{th}$ direction is non-compact, there are 
no winding modes.  We will take $p_\mu = 0$ for $\mu = 0,\ldots, 24$ 
because we are interested only in translationally invariant configurations.
Therefore, the string field is built by acting with the oscillators
$\alpha_n^\mu$, $c_n$ and $b_n$ 
on the zero-momentum ground states of strings stretching from
brane  $i$ to brane $j$.  We will denote the ground states 
with $|ij\rangle$ and normalize them so that 
$\langle ij | kl \rangle = \delta_{il}\delta_{jk}$
(i.e. $\left (|ij\rangle \right )^\dagger = \langle ji |$).

\label{hermicity}

Consider then a matrix-valued spacetime field $A$ of the string field, 
$\sum _{ij} A_{ij}{\cal{A}}|ij\rangle$,
 where $\cal A$ is an operator built out of $\alpha_n^\mu$, $c_n$ and $b_n$.
For the string field
to be real, it must be invariant under the combination of 
bpz and hermitian conjugation \cite{Gaberdiel:1997ia}.  Therefore, if $\beta_A$, defined by
$(bpz({\cal A}))^\dagger= \beta_A {\cal A}$, is $+1$,
the matrix of fields $A$ is hermitian, $A_{ij} = \overline{A_{ji}}$. 
If $\beta_A = -1$, $A$ must be anti-hermitian, $A_{ij} = -\overline{A_{ji}}$. 
We will refer to the first class of spacetime fields 
as `real' and the second class as `imaginary'.

To check that the quadratic part of the action, proportional to
$\left<\Phi|Q_B \Phi\right>$, is real under this hermicity condition, 
let the string field $|\Phi\rangle$ contain a term  $\sum _{ij} A_{ij}{\cal{A}}|ij\rangle$ and 
$|Q_B \Phi\rangle$ contain $\sum _{ji}B_{ji}{\cal{B}}|ji\rangle$,
with matrix-valued fields $A$ and $B$.
Now $\left<\Phi|Q_B\Phi\right>$ will contain two cross terms
\be
\sum_{ij} A_{ij} B_{ji} \left (
\beta_A \left < ij | {\cal{A}}^\dagger {\cal{B}} | ji \right > \right )
~+~
\sum_{ij} A_{ij} B_{ji} \left (
\beta_B \left < ji | {\cal{B}}^\dagger {\cal{A}} | ij \right > \right )
\label{quadratic-coefficient}
\ee
which, with $\beta_A A_{ij} = (A^\dagger)_{ij} = \overline {A_{ji}}$ and
$\beta_B B_{ji} = (B^\dagger)_{ji} = \overline {B_{ij}}$,
can be combined to give
\bear
&\sum_{ij}  \beta_A \left ( A_{ij} B_{ji}
\left < ij | {\cal{A}}^\dagger {\cal{B}} | ji \right >\right )
~+~
\sum_{ij}  \beta_A  \left ( \overline {A_{ji} B_{ij}}
\left < ji | {\cal{B}}^\dagger {\cal{A}} | ij \right >\right ) 
=& \nn \\  
&\sum_{ij} 2 \beta_A \mathrm{Re}\left ( A_{ij} B_{ji}
\left < ij | {\cal{A}}^\dagger {\cal{B}} | ji \right >\right )~,&
\eear
which is real as required.

Let us now check that the cubic part of the action is real as well.
The cubic part of the action
is proportional to $\left<\Phi|\Phi\ast\Phi\right>$ and can
be written in terms of the three-string vertex $\bra{V_{3}}$ as
\be
\label{eq.sep.3vertex-cft}
\bra{V_{3}}\ket{\Phi^{(1)}}\ket{\Phi^{(2)}}\ket{\Phi^{(3)}}=
\left<\Phi|\Phi\ast\Phi\right>.
\ee
Consider the twist symmetry, $\Omega$, 
which reverses the orientation of the string.
The three-string vertex is invariant under  $\Omega$. Notice that
the mode expansion in (\ref{dirichlet}) implies 
$\Omega\alpha^{25}_n\Omega^{-1} =
(-1)^{n+1} \alpha^{25}_n$ and, in particular, $\Omega(\alpha^{25}_0)
= - \alpha^{25}_0$.  This is consistent with the fact that $\Omega|ij\rangle
=-|ji\rangle$ (recall that the one-D-brane is twist odd, so that 
$\Omega|ii\rangle =-|ii\rangle$).  
Now, consider three terms of a string field: ${\cal{A}}|ij\rangle$,
${\cal{B}}|jk\rangle$, and ${\cal{C}}|ki\rangle$, such that
$\Omega{\cal W}\Omega^{-1} = \Omega_{W}{\cal W}$ 
for ${\cal W} = {\cal A},{\cal B},{\cal C}$.  
Define
\be
g(d_{ij}, d_{jk}, d_{ki}) ~\mathop{=}^{\mathrm{def}}~ \langle V_3| 
~{\cal{A}}|ij\rangle~ {\cal{B}}|jk\rangle ~{\cal{C}}|ki\rangle~.
\label{coeff-g}
\ee
Using the general properties of the string vertex, together with the 
fact that the string field is Grassmann odd and that the vacuum is twist odd,
it is easy to show that
\be
\langle V_3| ~ {\cal{C}}|ik\rangle~ {\cal{B}}|kj\rangle ~{\cal{A}}|ji\rangle
= - \Omega_A \Omega_B \Omega_C g(d_{ij}, d_{jk}, d_{ki})~.
\ee
Trivially, it is also true that
\be
\langle V_3| ~ {\cal{B}}|jk\rangle ~{\cal{C}}|ki\rangle~{\cal{A}}|ij\rangle~=
\langle V_3| ~{\cal{C}}|ki\rangle~{\cal{A}}|ij\rangle~ {\cal{B}}|jk\rangle ~
= g(d_{ij}, d_{jk}, d_{ki})~.
\ee
If the total string field contains these three terms with
spacetime fields $A_{ij}$, $B_{jk}$ and $C_{ki}$:
\be
|\Phi\rangle = \ldots + 
\sum _{ij} A_{ij}{\cal{A}}|ij\rangle 
+ \sum _{jk}B_{jk}{\cal{B}}|jk\rangle +
\sum _{ki} C_{ki}{\cal{C}}|ki\rangle + \ldots~,
\ee
then the cubic part of the action, 
$\langle \Phi | \Phi * \Phi\rangle$, contains a term
\be
3 \sum_{ijk} g(d_{ij}, d_{jk}, d_{ki}) \left ( A_{ij} B_{jk} C_{ki}  
- \Omega_A \Omega_B \Omega_C  C_{ik} B_{kj} A_{ji} \right )~.
\label{action.term}
\ee
Now, write all three fields $\cal A$, $\cal B$ and $\cal C$ in terms of
the creation operators $\alpha_{-n}$, $c_{-m}$, $b_{-k}$ ($n>0$, $m>-2$,
$k>1$).  Looking at the following summary of the behaviour of different
oscillators under the combined bpz and hermitian conjugation and under
twist:
\be
\begin{array}{rclcrcl}
\Omega \alpha_n \Omega^{-1} & = & (-1)^{n+1} \alpha_n &~~~~~&
(bpz(\alpha_n))^\dagger &=& (-1)^{n+1} \alpha_n \\
\Omega c_n \Omega^{-1} & = & (-1)^{n} c_n &~~~~~&
(bpz(c_n))^\dagger &=& (-1)^{n+1} c_n \\
\Omega b_n \Omega^{-1} & = & (-1)^{n} b_n &~~~~~&
(bpz(b_n))^\dagger &=& (-1)^{n} b_n 
\end{array}
\ee
we see that $\Omega_A = \beta_A (-1)^{N_c}(-1)^{N_F(N_F-1)/2}$ where 
$N_c$ is the number of $c_n$-oscillators in the field $\cal A$
and $N_F$ is the number of fermionic oscillators (the last
factor comes from the properties of the $bpz$ conjugate while
acting on a product of fermionic oscillators \cite{Taylor:2003gn}).  Since the field has ghost 
number 1, the number of $b_n$ oscillators must be $N_c-1$ and
therefore $N_F = 2N_c - 1$.  This implies that
$\Omega_A = - \beta_A$ (which means that spacetime fields are
hermitian (anti-hermitian) matrices if the twist eigenvalue of
associated state ${\cal A}$ is negative (positive)). We can rewrite equation 
(\ref{action.term}) as
\be
3 \sum_{ijk} g(d_{ij}, d_{jk}, d_{ki}) \left ( A_{ij} B_{jk} C_{ki}  
+ \beta_A \beta_B \beta_C  C_{ik} B_{kj} A_{ji} \right )~.
\ee
Using $\beta_A A_{ji} = (A^\dagger)_{ji} = \overline {A_{ij}}$ (and
similarly for $B$ and $C$) we see that the above expression is always
real, as required, and equal to
\be
6 \sum_{ijk} g(d_{ij}, d_{jk}, d_{ki}) 
\mathrm{Re}\left ( A_{ij} B_{jk} C_{ki} \right )~.
\ee

As is usual in tachyon condensation computations, we restrict ourselves
to twist even string fields.  However, since the 
twist $\Omega$ acts in a non-diagonal way on the  $|ij\rangle$ basis
($\Omega |ij\rangle = - |ji\rangle$), restricting to
twist even is not the same as restricting to even level fields.
To restrict our solutions to twist even fields, we can act on the string
field with $\half(1+\Omega)$.  In case of a real spacetime field corresponding
to an operator $\cal A$, with
$-\Omega_A = \beta_A = +1$ we obtain
\be
\half (1+\Omega) A_{ij} {\cal A} |ij\rangle = 
\half ( A_{ij} + A_{ji}) {\cal A} |ij\rangle
\ee
and therefore we can restrict ourselves to a matrix of fields $A$ which
is not only hermitian, but also symmetric and therefore has real entries.
In case of an imaginary spacetime field, $-\Omega_A = \beta_A = -1$ we 
we have
\be
\half (1+\Omega) A_{ij} {\cal A} |ij\rangle = 
\half ( A_{ij} - A_{ji}) {\cal A} |ij\rangle~,
\ee
which means we can restrict ourselves to an antihermitian antisymmetric matrix
$A$, which again has purely real entries.

In the twist even sector, we wish now to impose Siegel gauge.  
Since this gauge can be imposed level by level for the usual reasons, 
we can do so at all levels except for level one, where
$L_0=0$ and the usual argument for local validity of Siegel gauge fails.  
We therefore will impose Siegel gauge starting at level two, and 
will include the level one state $c_0|ij\rangle$.\footnote
{For separated D-branes, we could have imposed Siegel gauge
in the off-diagonal sector at level one, setting
$c_0|ij\rangle$ to zero as long as $i\neq j$, but we will not 
find this to be necessary, since for two D-branes, the exchange symmetry 
described in section \ref{2branes} will allow us to set the corresponding
fields to zero anyway.}

We are now ready to write down the real, twist even, (mostly) Siegel
gauge string field.  We can use rotational invariance in the 
directions parallel to the 
D24-branes to argue that $\alpha_{-n}^\mu$ oscillators for 
$\mu = 0, \ldots, 24$ can appear only in combinations of the form
$\alpha_{-n}\cdot\alpha_{-m}  = \sum_{\mu=0}^{24} 
\alpha_{-n}^\mu \alpha_{-m}^\mu$.  We single out the oscillators in the 
direction normal to the D24-branes by a superscript $25$.
$L^{25}_{-n}$ are matter Viarasoro generators for the CFT associated
with the field $X^{25}$,  while 
$L'_{-n}$ are matter Viarasoro generators for the CFT associated
with the fields $X^{\mu}$ (with central charge 25),
so that $L^{m}_{-n} = L^{25}_{-n} + L'_{-n}$.
With this notation, the total string field up to level three is given in 
Tables \ref{string-field-L} and \ref{string-field}.
Table \ref{string-field-L} uses the Viarasoro generators in the 
matter sector, while  Table \ref{string-field} uses the $\alpha_{-n}$ 
oscillators.  The relationship between the fields corresponding
to these two different sets of states is given at the bottom of Table \ref{string-field}.
We include the presentation in the $\alpha_{-n}$ basis because,
at some D-brane separations, the Viarasoro basis ceases to be complete.  
For example, at $d_{ij}=\pm 1/\sqrt 2$ the 
states
$L_{-2}^{25}c_1|ij\rangle$ and  $L_{-1}^{25}\alpha_{-1}^{25}c_1|ij\rangle$ 
are no longer linearly independent.  This degeneracy implies that
the relationship between the fields in these
two bases is singular at these special separations:
at $d_{ij}=\pm 1/\sqrt 2$, the relationship between 
$(f_{ij},w_{ij})$ and  $(\tilde f_{ij},\tilde w_{ij})$
takes on a singular form $\tilde w_{ij} = w_{ij}/2 + f_{ij}/\sqrt 2$, 
$\tilde f_{ij} = w_{ij}/\sqrt 2 + f_{ij}$,  and
$(f_{ij},w_{ij})$ becomes infinite while $(\tilde f_{ij},\tilde w_{ij})$
remains finite.  Since we will never use a D-brane separation which is 
exactly equal to one of these special values, we will
perform the computations in the Viarasoro-generated basis.
However, we will use some of the fields from 
the other basis to present our results, as these remain finite everywhere.

\begin{table}
\begin{center}
\renewcommand{\arraystretch}{1.6}
\begin{tabular}{|c|l|}
\hline  Level & String field $|\Phi\rangle$ \\ \hline \hline
0 & $t_{ij}c_{1}|ij\rangle$  \\ \hline
1 & $\left ( h_{ij}c_0 ~+ ~x_{ij} \alpha_{-1}^{25}  c_{1}\right ) |ij\rangle $
\\ \hline
2 & $\left ( u_{ij} c_{-1} ~+~ 
v_{ij} L'_{-2} c_1  ~+~
 w_{ij} L_{-2}^{25} c_1 ~+~ 
f_{ij} L_{-1}^{25}\alpha_{-1}^{25} c_{1}  
\right ) |ij \rangle $ \\ \hline
3 &  $\left (   (o_1)_{ij} b_{-2}c_{-1}c_1 + (o_2)_{ij} c_{-2} ~+~ 
r_{ij} \alpha_{-1}^{25}  c_{-1}~+~
s_{ij} L'_{-2} \alpha_{-1}^{25} c_1  ~+~
p_{ij} L'_{-3}  c_1  
\right . $ \\ ~ & $ ~~~\left . ~+~
q_{ij} L_{-3}^{25}  c_1  ~+~
y_{ij} L_{-2}^{25} \alpha_{-1}^{25} c_1 ~+~ 
z_{ij} L_{-1}^{25}L_{-1}^{25}\alpha_{-1}^{25}  c_1  
\right ) |ij \rangle $ \\ \hline
\end{tabular} 
\caption{Level three string field in (mostly) Siegel gauge.  $t$, $x$, $u$,
$v$, $w$, $r$, $s$, $y$, $z$
are real symmetric matrices while $h$, $f$, $o_1$, $o_2$
$p$, $q$ are real antisymmetric ones.}
\label{string-field-L}
\renewcommand{\arraystretch}{0.625}
\end{center}
\end{table}

\begin{table}
\begin{center}
\renewcommand{\arraystretch}{1.6}
\begin{tabular}{|c|l|}
\hline  Level & String field $|\Phi\rangle$\\ \hline \hline
0 & $t_{ij}c_{1}|ij\rangle$  \\ \hline
1 & $\left ( h_{ij}c_0 ~+ ~x_{ij} \alpha_{-1}^{25}  c_{1}\right ) |ij\rangle $
\\ \hline
2 & $\left ( u_{ij} c_{-1} ~+~ 
\tilde v_{ij} \alpha_{-1}\cdot \alpha_{-1} c_1  ~+~
\tilde w_{ij} \alpha_{-1}^{25} \alpha_{-1}^{25} c_1 ~+~ 
{\tilde f_{ij}} \alpha_{-2}^{25} c_{1}  
\right ) |ij \rangle $ \\ \hline
3 &  $\left (      (o_1)_{ij} b_{-2}c_{-1}c_1 + (o_2)_{ij} c_{-2} ~+~  
r_{ij} \alpha_{-1}^{25}  c_{-1}~+~
\tilde s_{ij} \alpha_{-1}\cdot \alpha_{-1} \alpha_{-1}^{25} c_1  ~+~
p_{ij} \alpha_{-1}\cdot \alpha_{-2}  c_1  
\right . $ \\ ~ & $ ~~~\left . 
~+~ \tilde q_{ij} \alpha_{-1}^{25} \alpha_{-2}^{25}  c_1  ~+~
\tilde y_{ij} \alpha_{-1}^{25} \alpha_{-1}^{25} \alpha_{-1}^{25} c_1 ~+~ 
{\tilde z}_{ij} \alpha_{-3}^{25}  c_1  
\right ) |ij \rangle $ \\ \hline
\end{tabular} 
$$\tilde v_{ij}  = \half v_{ij} ~~~~~~
 \tilde w_{ij}  = \half w_{ij} + f_{ij} d_{ij}~~~~~~
 \tilde f_{ij}= f_{ij} + w_{ij} d_{ij} ~~~~~~
 \tilde s_{ij} = \half s_{ij}  
$$
$$
\tilde q_{ij} =q_{ij} + y_{ij}d_{ij} + 3z_{ij}d_{ij} ~~~~~~
\tilde y_{ij} = \half y_{ij} + z_{ij} (d_{ij})^2  ~~~~~~
\tilde z_{ij} = 2 z_{ij} + y_{ij} + q_{ij} d_{ij}  
$$
\caption{Level three string field in (mostly) Siegel gauge, written using 
$\alpha_{-n}$ oscillators instead of $L_n$.  $t$, $x$, $u$,
$\tilde v$, $\tilde w$, $r$, $\tilde s$, $\tilde y$, $\tilde z$
are real symmetric matrices while $h$, ${\tilde f}$, $o_1$, $o_2$
$p$, $\tilde q$ are real antisymmetric ones.}
\label{string-field}
\renewcommand{\arraystretch}{0.625}
\end{center}
\end{table}

\subsection{Level truncation}
\label{truncation}

Following \cite{Moeller:2000jy}, we should define the level of the string field
to include the total $L_0$ eigenvalue, and not just that part of it which
counts oscillator excitations.  Specifically, for a string stretched
between D-branes $i$ and $j$, we have
\be
L_0 = \half (d_{ij} )^2 + \tilde N~,
\ee
where $\tilde N$ is the contribution to $L_0$ from the non-zero
matter and the ghost oscillators.
When acting on the diagonal part of the tachyon field, $L_0$ gives 
$L_0 c_{1} |ii \rangle = - c_{1}|ii \rangle $, and therefore we should 
perhaps define the level to be
\be
l = \half (d_{ij} )^2 + (\tilde N+1)
\ee
Then, once $d_{ij}$ is large enough, we would need to include fields with higher 
$\tilde N$ before including off-diagonal
fields with lower $\tilde N$.  For example, with two D-branes, 
diagonal fields with $\tilde N=4$ should be included before
off-diagonal fields with $\tilde N=3$ if the D-brane separation $d_{12}$ 
is greater than $\sqrt 2$.  However, since we will be studying
the solutions as functions of the D-brane separation, changing which fields are 
included would cause discontinuities, making our results
hard to interpret.  We will therefore keep the field content of our
truncated string field the same at all separations and, for simplicity,
we will refer to $(\tilde N+1)$ as the level.  However, we should keep in 
mind that, for two D-branes, potentially more accurate results could be obtained
for $d_{ij}>\sqrt{2}$ by adapting the field content accordingly.

\subsection{The action}
\label{action}

The OSFT potential (in units of 24-dimensional D-brane tension) can be expressed as
\be \label{eq.sep.basic-action}
f(\ket{\Phi}) = -{S(\ket{\Phi}) \over M}
=2\pi^{2}\left(\frac{1}{2}\langle\Phi,Q_{B}  \Phi\rangle
+\frac{1}{3}\bra{V_{3}}\ket{\Phi^{(1)}}\ket{\Phi^{(2)}}\ket{\Phi^{(3)}}\right)
\ee
where the BRST operator is
\be
Q_{B}=c_{n}L_{-n}^{m}~+~\frac{m-n}{2}:c_{m}c_{n}b_{-m-n}:~-~c_{0},
\ee
while the three-string vertex $\bra{V_{3}}$ was defined in equation 
(\ref{eq.sep.3vertex-cft}) and can be written as
\be \label{eq.sep.vertex-defn}
\bra{V_{3}}=\frac{3^{4}\sqrt{3}}{2^{6}}\sum_{i,j,k}\bra{ij}\bra{jk}\bra{ki}
	c_{-1}^{(1)}c_{-1}^{(2)}c_{-1}^{(3)}c_{0}^{(1)}c_{0}^{(2)}c_{0}^{(3)}
	e^{\Xi}~,
\ee
with
\be
\Xi=\sum_{r,s=1}^3\sum_{m,n=0}^{\infty}\left(\frac{1}{2}\alpha_{m}^{(r)\mu}N_{mn}^{rs}\alpha_{n,\mu}^{(s)} \right )
~+~
\sum_{r,s=1}^3\sum_{m=1,n=0}^{\infty}\left(c_{m}^{(r)}X_{mn}^{rs}b_{n}^{(s)}\right)~.
\label{eq.sep.xi-defn} 
\ee
As usual, the $r,s$ indices run from 1 to 3 and 
label the three strings interacting at the vertex.

To compute the quadratic terms in the potential, we use the 
properties of the $bpz$ conjugate,
\be
\left|ij\right>\rightarrow\left<ij\right|,\quad
\alpha^\mu_{n}\rightarrow(-1)^{n+1}\alpha^\mu_{-n},\quad
L_{n}\rightarrow(-1)^{n}L_{-n}
\ee
\be
c_{n}\rightarrow(-1)^{n+1}c_{-n},\quad
b_{n}\rightarrow(-1)^{n}b_{-n},\quad
bpz(AB)=bpz(A)bpz(B)
~\footnote{For purely commuting or purely anti-commuting A and B, $AB = \pm BA$
\cite{Taylor:2003gn}.}~.
\ee

To compute the cubic terms in the potential, we need the three-string 
coefficients  $N_{mn}^{rs}$ and $X_{mn}^{rs}$.  These are well 
known \cite{Gross:1986fk} and were originally computed for Neumann strings.
T-duality guarantees that the same result applies to Dirichlet boundary
conditions, at least as long as the D-brane coordinates are diagonal.
The three-string coefficients we require are given in Table 
\ref{tab.sep.neumann-coeffs}.
It will be useful later to define $\Nt^{rs}_{mn}=N^{rs}_{mn}+N^{sr}_{nm}$.  
Given the string field $\Phi$, it is straight-forward to compute 
$e^{\Xi}\Phi e^{-\Xi}$.  Using the fact that 
\be 
e^{\Xi}|ij\rangle|jk\rangle|ki\rangle = 
\left ( 4\over 3\sqrt 3\right )^{\half((d_{ij})^2 + (d_{jk})^2 + (d_{ki})^2)
}|ij\rangle|jk\rangle|ki\rangle~,
\ee
the coefficients $g(d_{ij},d_{jk},d_{ki})$ defined in equation (\ref{coeff-g})
can then be obtained using a computer-assisted algebra program.

\begin{table}
\begin{center}
\begin{tabular*}{0.75\textwidth}{@{\extracolsep{\fill}}|c|ccc|}
	\hline
	m~~n & $N_{mn}^{rr}$ & $N_{mn}^{r(r+1)}$ & $N_{mn}^{r(r-1)}$ \\
	\hline \hline
	0~~0& $\ln 4\sqrt{3}/9$ & 0 & 0 \\
	\hline
	0~~1& 0 & $-2\sqrt{3}/9$ & $2\sqrt{3}/9$ \\
	1~~0& 0 & $2\sqrt{3}/9$ & $-2\sqrt{3}/9$ \\
	\hline
	0~~2& $-2/27$ & $1/27$ & $1/27$ \\
	1~~1& $-5/27$ & $16/27$ & $16/27$ \\
	2~~0& $-2/27$ & $1/27$ & $1/27$ \\
	\hline
	0~~3& 0 & $22\sqrt{3}/729$ & $-22\sqrt{3}/729$ \\
	1~~2& 0 & $32\sqrt{3}/243$ & $-32\sqrt{3}/243$ \\
	2~~1& 0 & $-32\sqrt{3}/243$ & $32\sqrt{3}/243$ \\
	3~~0& 0 & $-22\sqrt{3}/729$ & $22\sqrt{3}/729$ \\
	\hline
\end{tabular*}
\vskip 0.2in
\begin{tabular*}{0.75\textwidth}{@{\extracolsep{\fill}}|c|ccc|}
	\hline
	m~~n & $X_{mn}^{rr}$ & $X_{mn}^{r(r+1)}$ & $X_{mn}^{r(r-1)}$ \\
	\hline \hline
	1~~0& 0 & $4\sqrt{3}/9$ & $-4\sqrt{3}/9$ \\
	\hline
	1~~1& $11/27$ & $8/27$ & $8/27$ \\
	2~~0& $16/27$ & $-8/27$ & $-8/27$ \\
	\hline
	1~~2& 0 & $40\sqrt{3}/243$ & $-40\sqrt{3}/243$ \\
	2~~1& 0 & $-80\sqrt{3}/243$ & $80\sqrt{3}/243$ \\
	3~~0& 0 & $-68\sqrt{3}/243$ & $68\sqrt{3}/243$ \\
	\hline
\end{tabular*} 
\caption{The first few Neumann coefficients appearing in equation
(\ref{eq.sep.xi-defn}).  The matter coefficients are taken from
\cite{Rastelli:2001jb} and the ghost coefficients from \cite{Gross:1986fk}.
The non-zero mode matter coefficients and the ghost coefficients agree
with equations (141) and (147) in \cite{Taylor:2003gn}.}
\label{tab.sep.neumann-coeffs}
\end{center}
\end{table}

For example, at level (1,3), we obtain

\begin{multline}
f(|\Phi\rangle) = 
2\pi^{2}\left[\frac{1}{2} \sum_{ij}\left(\left(-1+\frac{(d_{ij})^2}{2}\right)t_{ij}t_{ji}
	+\frac{(d_{ij})^2}{2}x_{ij}x_{ji}-2h_{ij}h_{ji}\right)\right. \\
	+\frac{3^{3}\sqrt{3}}{2^{6}} \sum_{ijk}\left(t_{ij}t_{jk}t_{ki}
	+\frac{3}{2}\Nt^{r3}_{01}\al^{(r)}t_{ij}t_{jk}x_{ki}
	+\frac{3}{2}\left(\Nt^{32}_{11}+\frac{1}{2}\Nt^{r2}_{01}\Nt^{s3}_{01}\al^{(r)}\al^{(s)}\right)t_{ij}x_{jk}x_{ki}\right.\\
	+\frac{1}{4}\left(\Nt^{12}_{11}\Nt^{3r}_{10}\al^{(r)}+\Nt^{23}_{11}\Nt^{1r}_{10}\al^{(r)}+\Nt^{31}_{11}\Nt^{2r}_{10}\al^{(r)}
	+\frac{1}{2}\Nt^{r1}_{01}\Nt^{s2}_{01}\Nt^{t3}_{01}\al^{(r)}\al^{(s)}\al^{(t)}\right)x_{ij}x_{jk}x_{ki}\\
	\left.\left.+\frac{16}{9}t_{ij}h_{jk}h_{ki}+\frac{8}{9}\Nt^{r3}_{01}\al^{(r)}h_{ij}h_{jk}x_{ki}
	\right)\left(\frac{4}{3\sqrt{3}}\right)^{\frac{1}{2}((d_{ij})^2+(d_{jk})^2+(d_{ki})^2)}\right]
\end{multline}
where we have omitted the index $25$ on the $\al$ oscillators for clarity and where $\al^{(1)} \rightarrow d_{ij}$,
$ \al^{(2)}\rightarrow d_{jk}$ and $\al^{(3)} \rightarrow d_{ki}$.

\subsection{Two D-branes and exchange symmetry}
\label{2branes}

The results up to this point were applicable to any number $N$ of parallel D24-branes.
Specializing to $N=2$, we define $d = d_{21}$,
so that $\alpha^{25}_0 |21\rangle= d |21\rangle$ and 
$\alpha^{25}_0 |12\rangle= -d |12\rangle$. $d$ is proportional to
the distance between the two D-branes.

A useful parametrization of the matrix-valued fields
(assuming a twist-even string field) is 
\be
t_{ij} = \left [ 
\begin{array}{cc} T_s - T_a &\tau \\ \tau & T_s+T_a\end{array}
 \right ] ~~~~~~~
x_{ij} = \left [ 
\begin{array}{cc} X_s - X_a &\chi \\ \chi & X_s+X_a\end{array}
 \right ]
\ee
and similar for all the real (and therefore symmetric) fields, and
\be
h_{ij}  = \left [ 
\begin{array}{cc} 0 &\gamma \\ -\gamma & 0\end{array}
 \right ]
\ee
and similar for all the imaginary fields.

At level (1,3) the potential is given explicitly by
\begin{multline} \label{eq.2brane.l1.potential}
f(T_s, T_a, \tau, X_s, X_a, \chi, \gamma)
 =2\pi^{2}\left[-T_{s}^{2}-T_{a}^{2}+\left(-1+\frac{d^{2}}{2}\right)\tau^{2}
	+\frac{d^{2}}{2}\chi^{2} + 2 \gamma^2\right.\\
+\frac{27\sqrt{3}}{32}T_{s}^{3}+\frac{81\sqrt{3}}{32}T_{s}T_{a}^{2}
	+\frac{3\sqrt{3}}{2}T_{s}(X_{s}^{2}+X_{a}^{2})
	+3\sqrt{3}T_{a}(X_{s}X_{a}) \\
+\left(\frac{4}{3\sqrt{3}}\right)^{d^{2}}\left(
	\frac{81\sqrt{3}}{32}T_{s}\tau^{2}+\frac{27 d}{8}X_{a}\tau^{2}
	-\frac{27 d}{8}T_{a}\tau\chi 
	+3\sqrt{3}\left(1-\frac{d^{2}}{2}\right)X_{s}\tau\chi
\right.\\ \left.\left.
	+\frac{3\sqrt{3}}{2}\left(1+\frac{d^{2}}{4}\right)T_{s}\chi^{2}
	+\frac{d^{3}}{2}X_{a}\chi^{2}-\frac{3\sqrt 3}{2}T_s\gamma^2
        -2d X_a\gamma^2 
\right)\right]
\end{multline}

When working with just two D-branes, there is an extra symmetry we can take
advantage of.  Since we are not interested in the marginal deformation which
moves both D-branes together in a rigid way, we will confine ourselves to
those solutions which correspond to  D-brane  configurations  
symmetric under $X^{25} \rightarrow -X^{25}$.  The action is invariant under 
simultaneously taking  $X^{25} \rightarrow -X^{25}$ (or, equivalently, $\alpha_n^{25} 
\rightarrow - \alpha_n^{25}$ for all $n$) and 
$1 \leftrightarrow 2$. We will refer to this as the exchange
symmetry.  If we are interested in solutions where the tachyon
field does not break this symmetry, we can restrict ourselves to
exchange symmetry even fields (as exchange symmetry odd fields cannot
appear linearly in the action).  This means that for fields with 
an even number of $\alpha_{-n}^{25}$ in them, we can set $A_{11}-A_{22}$
and $A_{12}-A_{21}$ to zero (this, at level one, implies that $T_a=0$
and $\gamma=0$) and for fields with an odd number of $\alpha_{-n}^{25}$, 
we can set $A_{11}+A_{22}$ and $A_{12}+A_{21}$ to zero (at level one, 
$X_s = 0$ and $\chi=0$).  It is easy to see from
the potential in equation (\ref{eq.2brane.l1.potential}) that we obtain
 a consistent
truncation of the theory at this level.  Truncating to exchange even
fields  allows us to drop $h$, $o_1$, $o_2$, $p$ and 
$q$ altogether.  With this truncation, we can compute
the complete cubic action at level (3,9).  The quadratic part of the potential up to level 3
can be found in the Appendix, while the cubic couplings including 
fields up to at level 3 will be published separately \cite{Longton:2012ei}. 
Where possible, the coefficients have been verified to agree with
those previously computed, for example in  \cite{Moeller:2000jy}.

\section{D-brane decay}

\subsection{Level 0}

\begin{figure}
\subfloat[] 
{\includegraphics[width=0.49\textwidth]{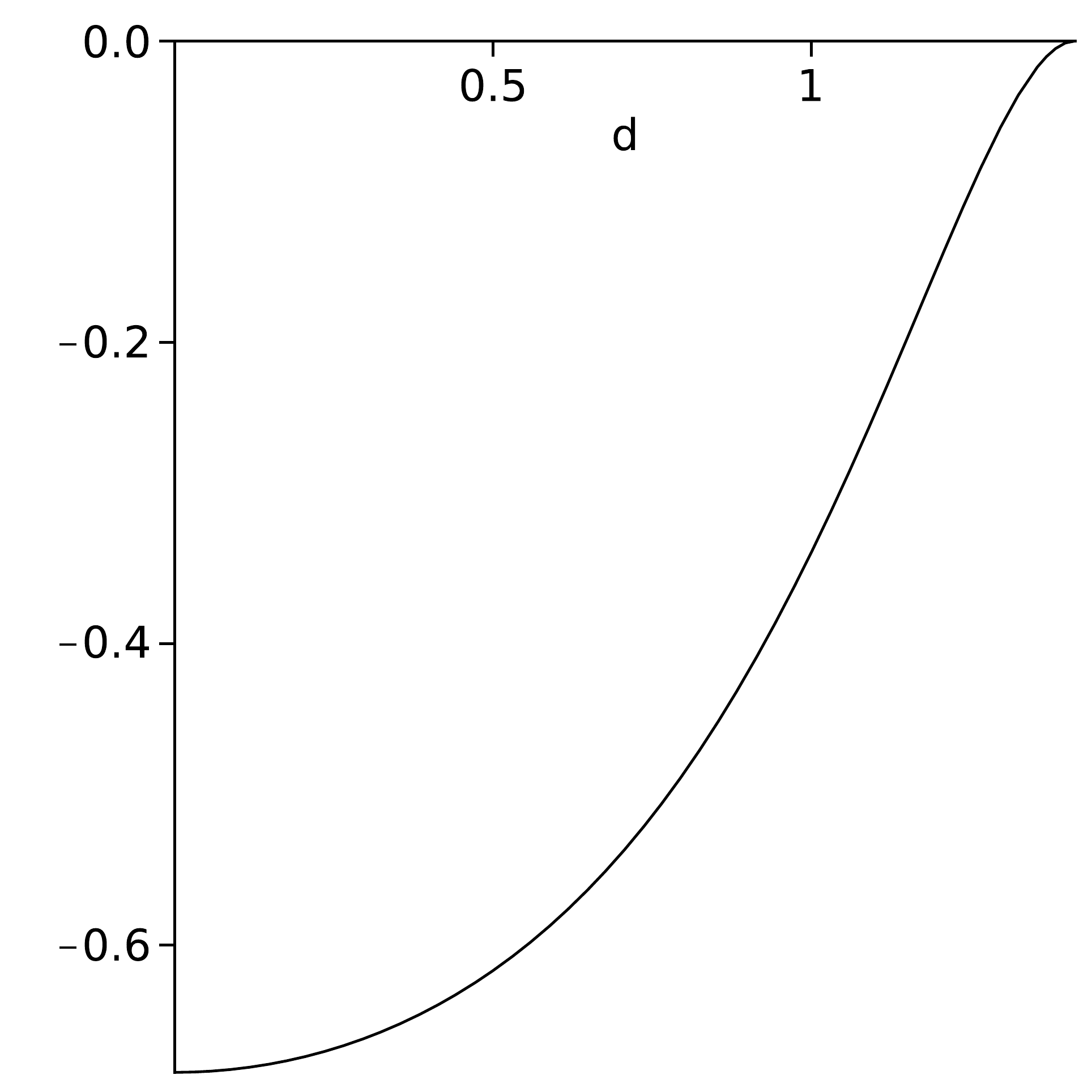}\label{f:level0a}}
\subfloat[]
{\includegraphics[width=0.49\textwidth]{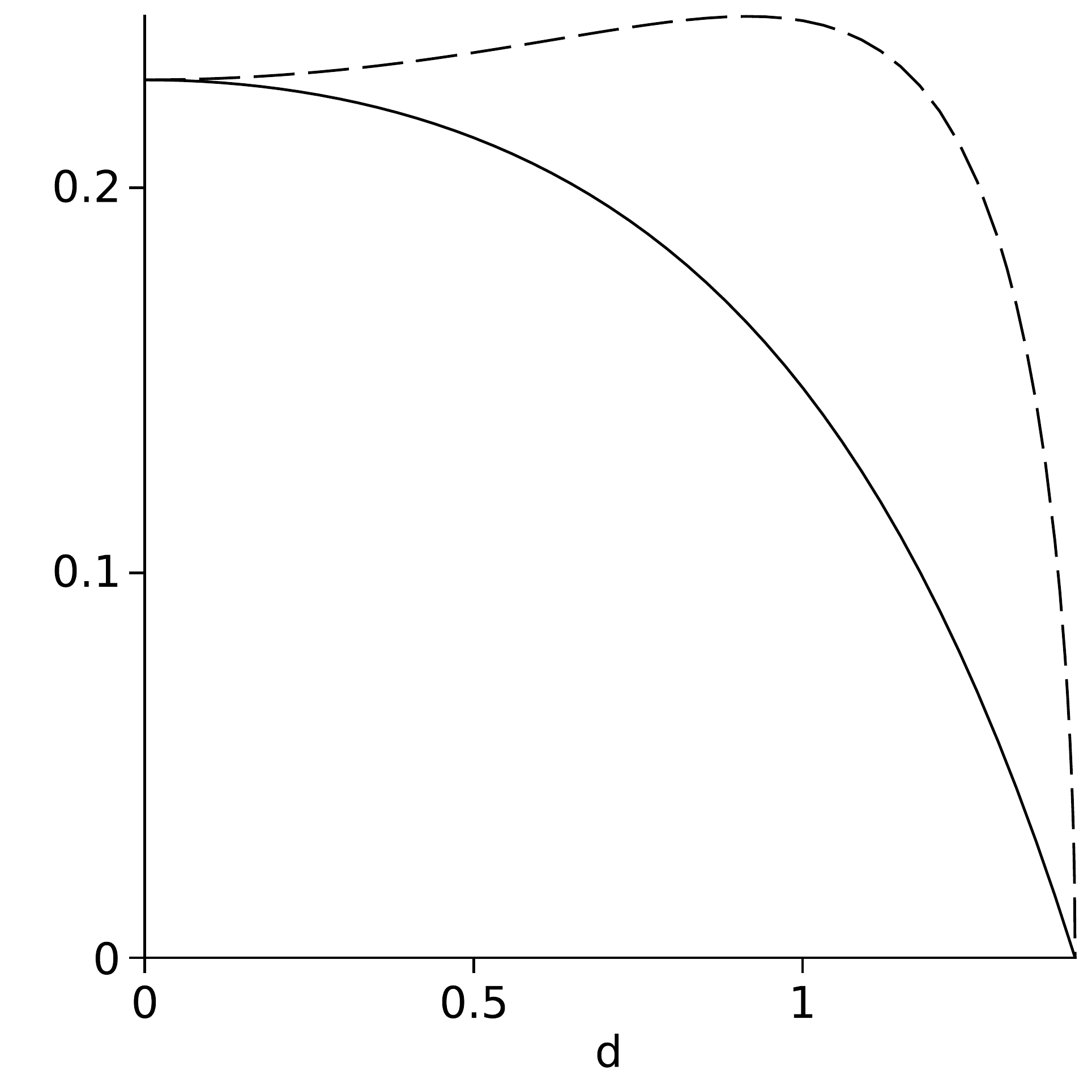}\label{f:level0b}}
\caption{The non-diagonal solution at level 0.  (a) Energy as a
function of D-brane separation, $d$. (b) The tachyon field:
$T_s$ (solid line) and $\tau$ (dashed line) as a function of D-brane
separation ($T_a=0$).}
\label{f:level0}
\end{figure}

We begin to analyze D-brane decay for two separated D-branes
at level 0, where only the tachyon fields come into play and where
the solutions can be obtained analytically.
The potential, without restricting to exchange even fields, is
\bear \label{eq.2brane.l0.potential}
&&f(T_s, T_a, \tau) = \\ \nn&&
 2\pi^{2}\left[-T_{s}^{2}-T_{a}^{2}+\left(-1+\frac{d^{2}}{2}\right)\tau^{2}
+\frac{81\sqrt{3}}{32}\left (
{1\over 3}T_{s}^{3}+ T_{s}T_{a}^{2}
+  \left(\frac{4}{3\sqrt{3}}\right)^{d^{2}} T_{s}\tau^{2}
\right)\right]~.
\eear
There are five  points where the derivatives
of the potential (\ref{eq.2brane.l0.potential}) vanish.  These include
the four expected solutions: the perturbative vacuum at $T_s=T_a=\tau=0$,
two solutions corresponding to one of the two D-branes decaying
($T_s=\pm T_a=T_0/2$, $\tau$=0), and one solution corresponding
to both D-branes decaying, ($T_s=T_0$, $T_a=0$, $\tau=0$), where
$T_0=64/81\sqrt 3$ is the well-know level 0 approximation to
the tachyon field (see for example \cite{Sen:1999nx}\cite{Taylor:2003gn}). 
In addition, there is a new, non-diagonal, exchange-even solution.
The energy and tachyon fields for this solution are shown in Figure 
\ref{f:level0}.  At zero separation, this solution corresponds simply
to single D-brane tachyon condensation, though the D-brane which undergoes decay
is an SU(2) rotation of our two original D-branes:
\bear
&&
\left [ 
\begin{array}{cc} T_s - T_a &\tau \\ \tau & T_s+T_a\end{array}
 \right ] = 
\left [ 
\begin{array}{cc} T_0/2 & T_0/2 \\ T_0/2 & T_0/2\end{array}
 \right ]  = 
U^\dagger \left [ \begin{array}{cc} T_0 & 0 \\ 0 & 0\end{array}
 \right ] U~, \\&& \nn \mathrm{where}~U = 
\left [ 
\begin{array}{cc} 1/\sqrt{2} & 1/\sqrt{2} \\ 1/\sqrt{2} & -1/\sqrt{2}\end{array}
 \right ]   ~.
\eear
The persistence of this solution for separated
D-branes, away from the SU(2) symmetric point, is at first surprisings.
Since it seems unlikely that this solution somehow corresponds to each of the
two D-branes having decayed `half-way', we propose the following
interpretation:  the solution corresponds to
the two separated D-branes moving towards each other
until coincident, restoring SU(2) symmetry before the decay occurs.  
The solution shown in Figure \ref{f:level0} should then correspond to
a combination of D-brane translation (a marginal deformation) and
tachyon decay.  To test this hypothesis, we analyze this solution at higher
truncation levels.  Since the solution is exchange-even at level 0, we
focus only on the exchange-even sector, as described 
in Section \ref{2branes}.

\subsection{Level (1,3)}

When we impose the exchange symmetry, the potential is quite simple
and involves only $T_s$, $\tau$ and $X_a$:  
\bear
f(T_s, \tau, X_a)
 =2\pi^{2}\left[ -T_{s}^{2}+\left(-1+\frac{d^{2}}{2}\right)\tau^{2} 
+\frac{27\sqrt{3}}{32}T_{s}^{3} 	+\frac{3\sqrt{3}}{2}T_{s}X_{a}^{2}
+\right . \\ \nn 
+ \left .
\left (\frac{4}{3\sqrt{3}}\right)^{d^{2}}\left(
	\frac{81\sqrt{3}}{32}T_{s}\tau^{2}+\frac{27 d}{8}X_{a}\tau^{2}
\right)\right]~,
\label{eq.2brane.l1.potential.with.exchangesym}
\eear
At this level, the string field includes the transverse scalar field $X_a$, which will
play a central point in the remainder of this paper.
This field is the infinitesimal marginal parameter for a D-brane 
translation mode moving the two D-branes symmetrically either
further apart or closer together \cite{Sen:2000hx}.

\begin{figure}
\subfloat[] 
{\includegraphics[width=0.3\textwidth]{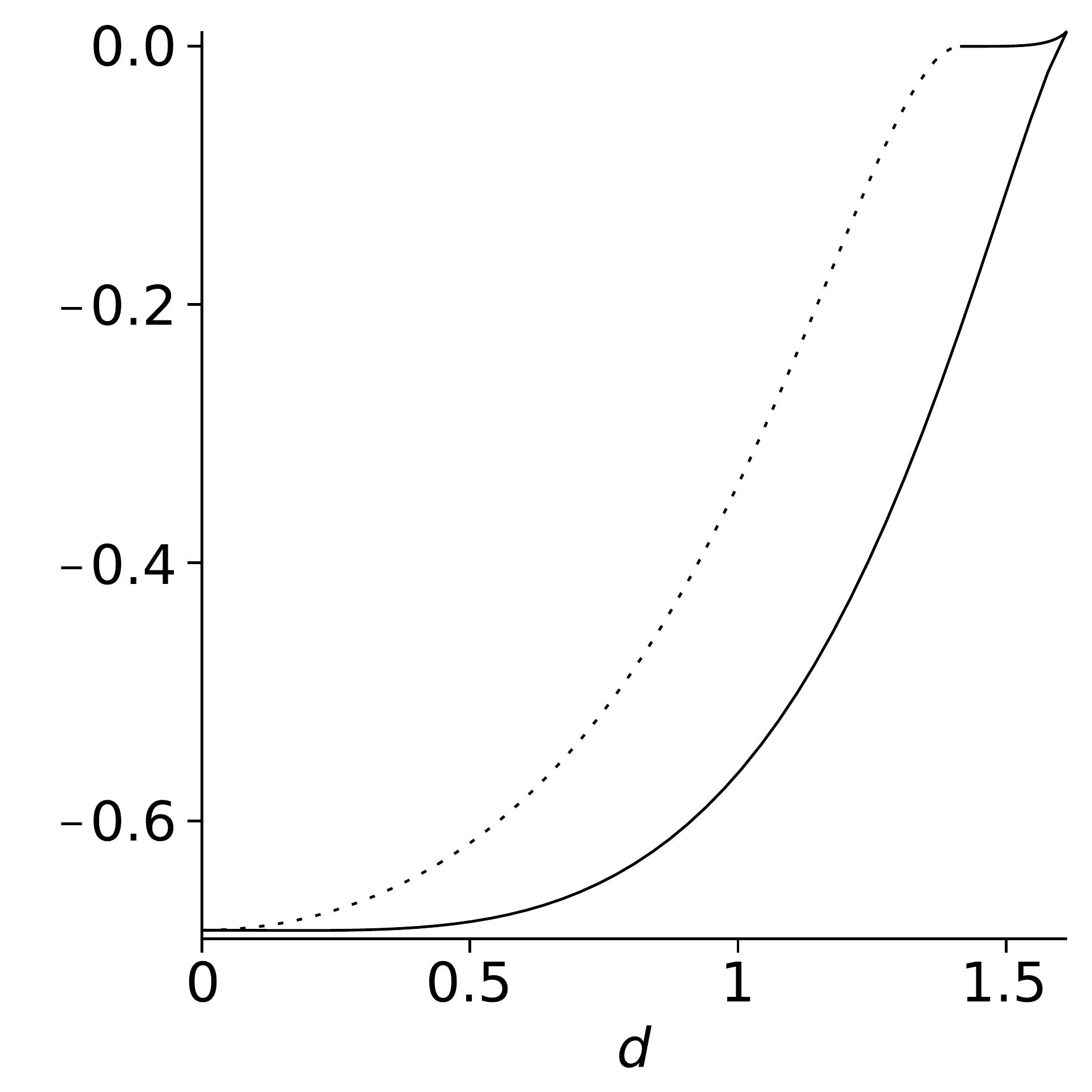}\label{f:level1a}} 
\subfloat[]
{\includegraphics[width=0.3\textwidth]{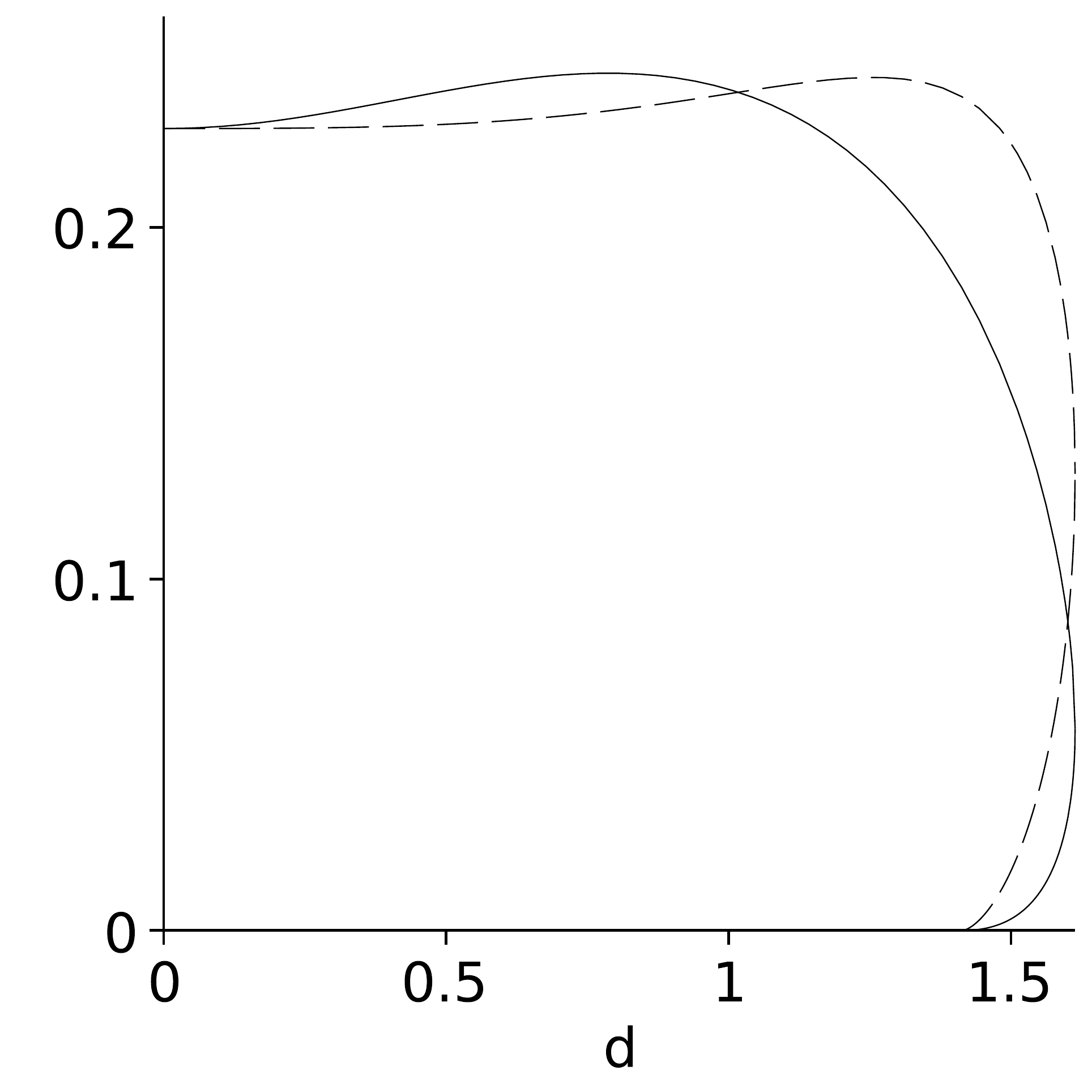}\label{f:level1b}}
\subfloat[]
{\includegraphics[width=0.3\textwidth]{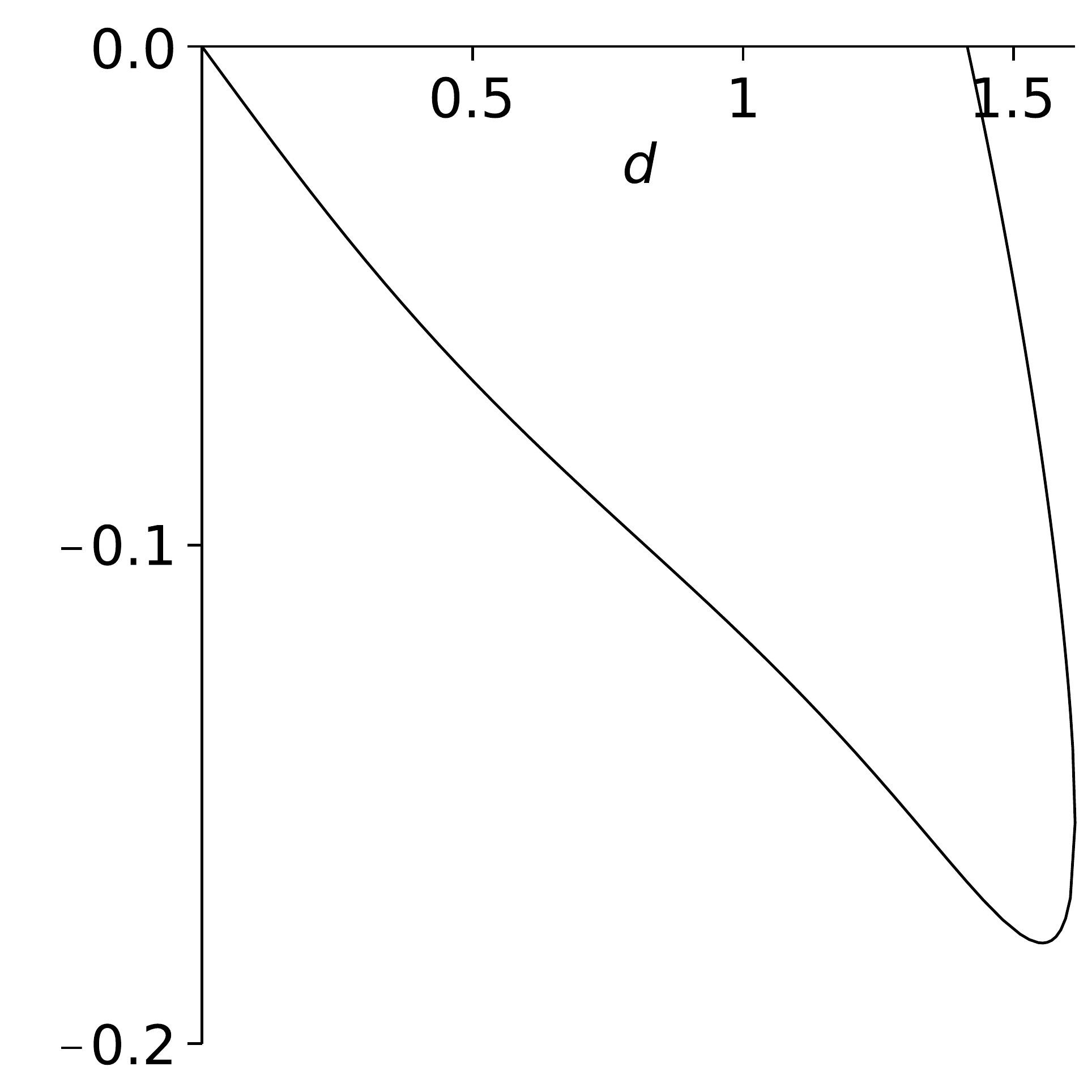}\label{f:level1c}}
\caption{The non-diagonal solution at level (1,3), 
as a function of initial D-brane separation, $d$.  (a) Energy  
(the dashed line shows the energy at level 0, from Figure 
\ref{f:level0a}, for comparison). 
(b) The tachyon field: $T_s$ (solid line) and $\tau$ (dashed line).  
(c) The transverse scalar field $X_a$.}
\label{f:level1}
\end{figure}

The non-diagonal solution of interest at this level is presented it Figure \ref{f:level1}.
It can be found analytically.

Let us first discuss several properties of this solution at relatively 
small values of D-brane separation.  
\begin{itemize}
\item $X_a$ is negative and approximately
linear as a function of separation.  The small $d$ behaviour in 
Figure \ref{f:level1c}  implies that the D-branes are moving
closer together by an amount proportional to the initial separation.
If, as we argued, this amount is in fact equal to the initial separation
(so that SU(2) symmetry is restored), Figure \ref{f:level1}c
can be interpreted as showing the relationship between the 
parameter $X_a$ and the physical distance by which the D-brane has moved
in our solution.
We will return to this point in section \ref{decay:discussion}.
\item  The energy as a function of separation is much flatter for 
level 1 than for level 0.  This again supports our hypothesis: 
if the solution we are studying corresponds to the two D-branes
coming together by a marginal deformation followed by a decay of 
some combination of the two coincident D-branes once SU(2) 
symmetry has been restored, the energy released in the decay should
be the same no matter what the original separation was.
\item As we already discussed, at zero separation, 
the solution has $T_s=\tau=T_0/2$, which is equivalent under
a SU(2) conjugation to $T_s \pm T_a = T_0$, $T_s \mp T_a  =\tau= 0$, 
the solution corresponding to the decay of either the left or the right brane.
As we see in Figure \ref{f:level1b}, $T_s \approx \tau \approx T_0/2$ up to
$d \approx 1.2$.  We take this to indicate that
at this level, the contribution to $T_s$ due to nonzero marginal
deformation parameter $X_a$ is quite small.  As we will see, this is
not so at higher levels.
\end{itemize}

Behaviour at larger separations is also quite interesting: at 
separation $d=\sqrt 2$, a new solution appears, and eventually
merges with our original solution around $d \sim 1.6$ (this is why
the plots in Figure \ref{f:level1} are double valued in this range).
$d = \sqrt 2$ is the point where the off-diagonal tachyon string
becomes massless.  It is not surprising that an appearance of a nearly
massless mode results in a new branch of the solution.  
That this branch merges with our solution of interest is somewhat similar
to what was found in \cite{Sen:2000hx}.
(This point will also be discussed in more detail 
in Section \ref{decay:discussion}.) 
No solutions can be found beyond the point where the two branches merge.
We also note that the branch which we interpret to represent tachyon decay
exists past the point where the off-diagonal tachyon string
becomes massless.  This further reinforces our interpretation of 
this branch: once the two D-branes have come together, the off-diagonal
tachyon is tachyonic once more.

\subsection{Level (3,9)}

\begin{figure}
\subfloat[] 
{\includegraphics[width=0.5\textwidth]{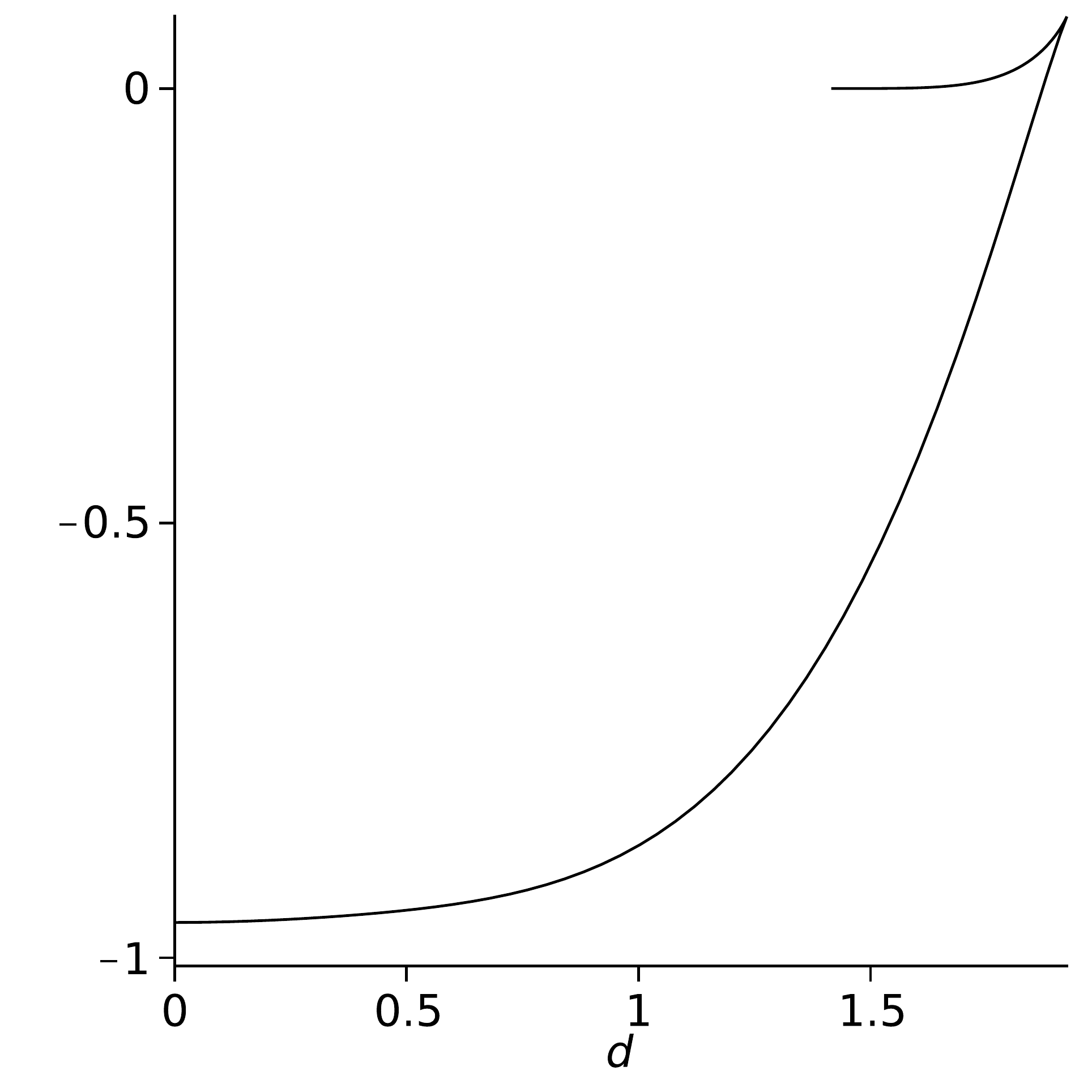}\label{f:level3a}} 
\subfloat[] 
{\includegraphics[width=0.5\textwidth]{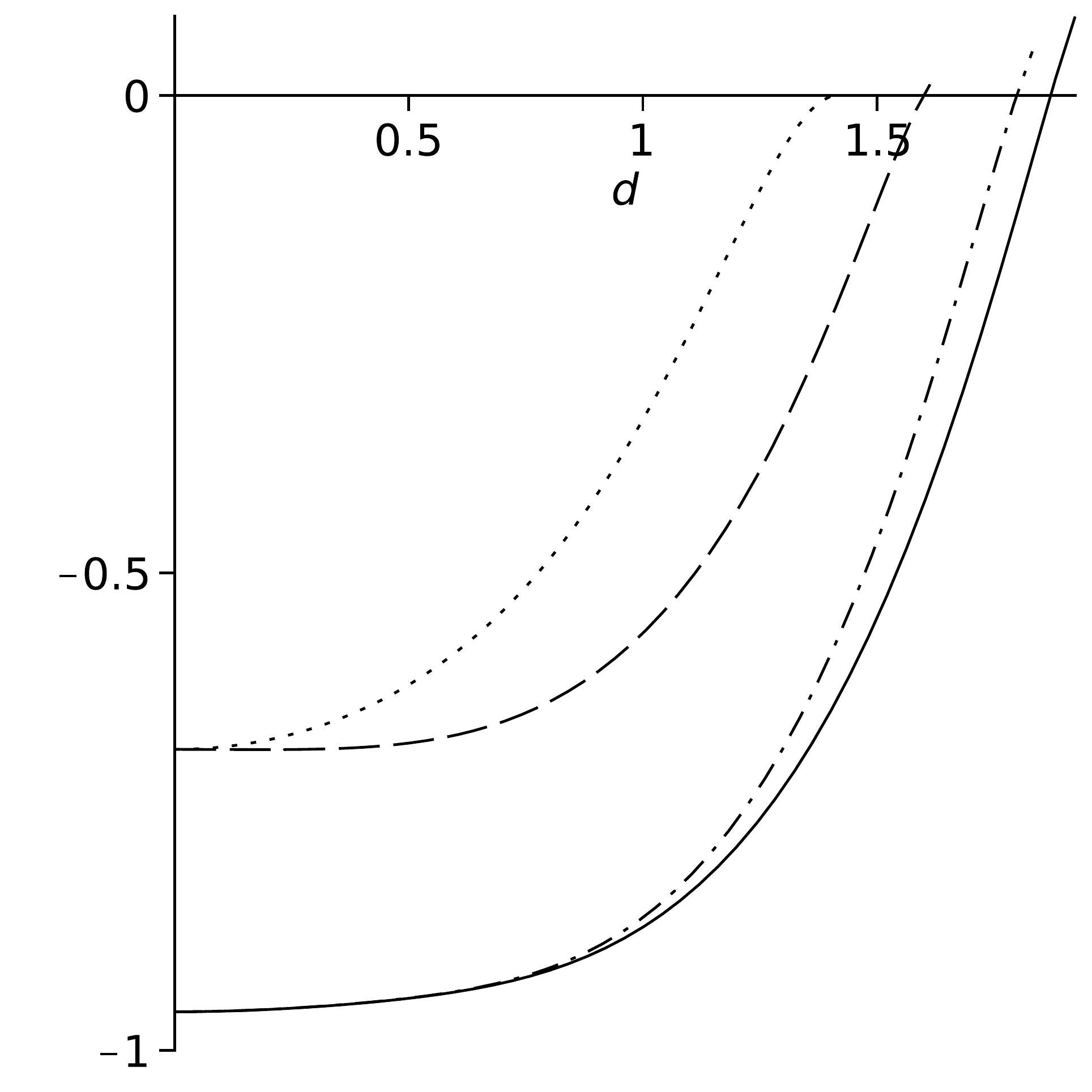}\label{f:level3b}} \\
\subfloat[]
{\includegraphics[width=0.5\textwidth]{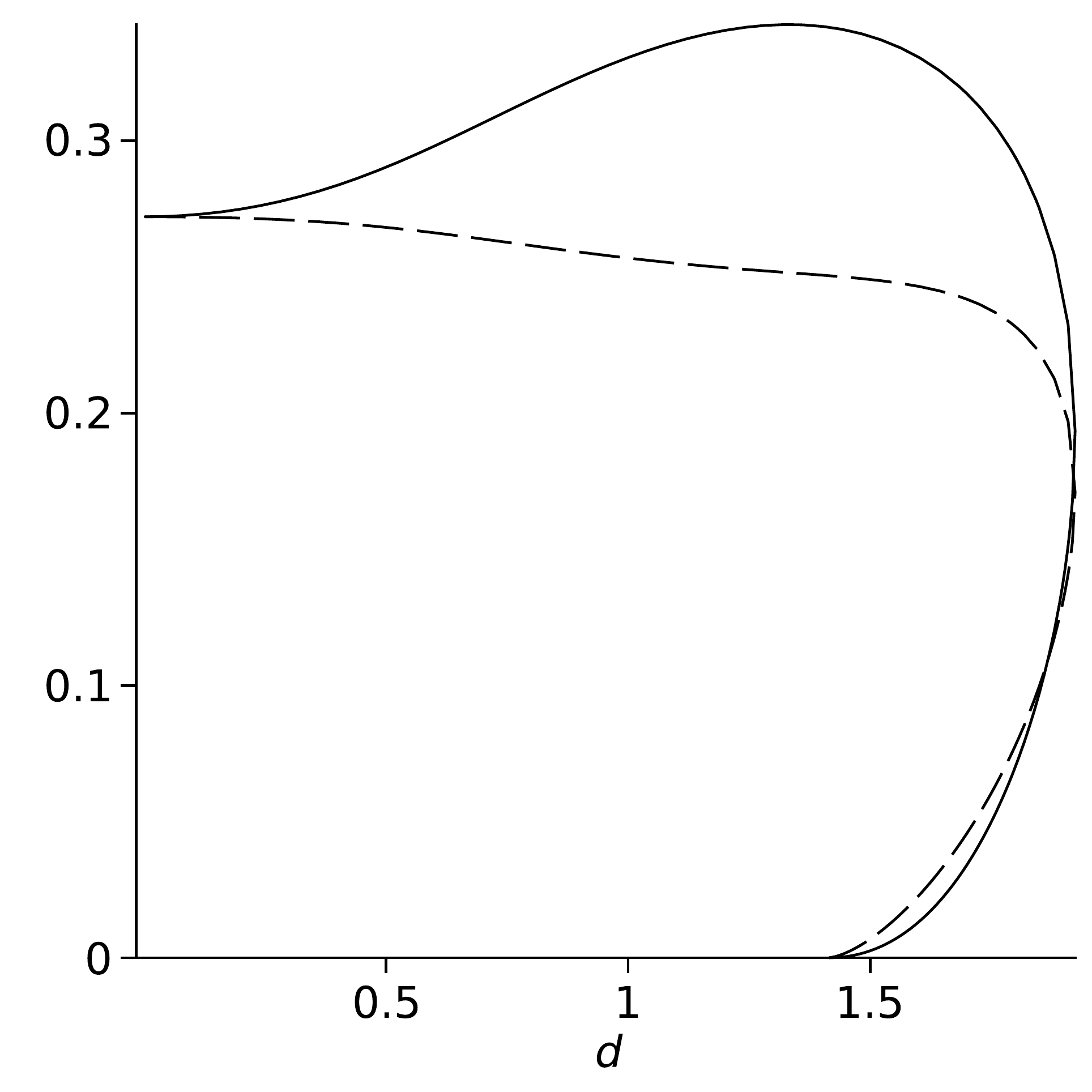}\label{f:level3c}} 
\subfloat[]
{\includegraphics[width=0.5\textwidth]{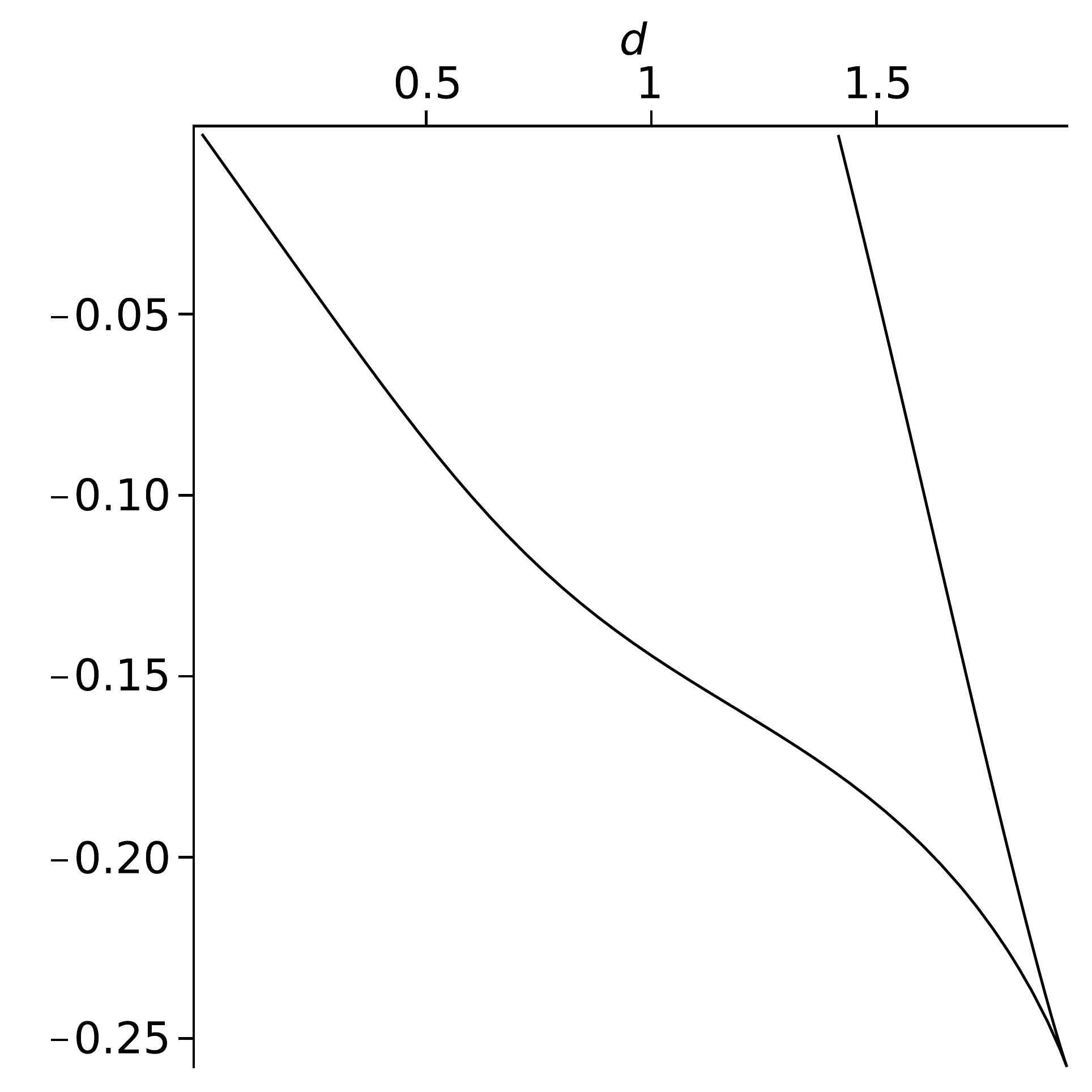}\label{f:level3d}} 
\caption{The non-diagonal solution at level (3,9), as a function of initial
D-brane  separation, $d$.  (a) Energy. (b) Energy at level 3 (solid line),
level 2 (dash-dot), level 1 (dash) and level 0 (dotted line),
for comparison.  For clarity, only the first branch is shown.  
(c) The tachyon field: $T_s$ (solid line) and $\tau$ (dashed line)
(d) The transverse scalar field $X_a$.}
\label{f:level3}
\end{figure}

\begin{figure}
	\subfloat[]
	{\includegraphics[width=0.23\textwidth]{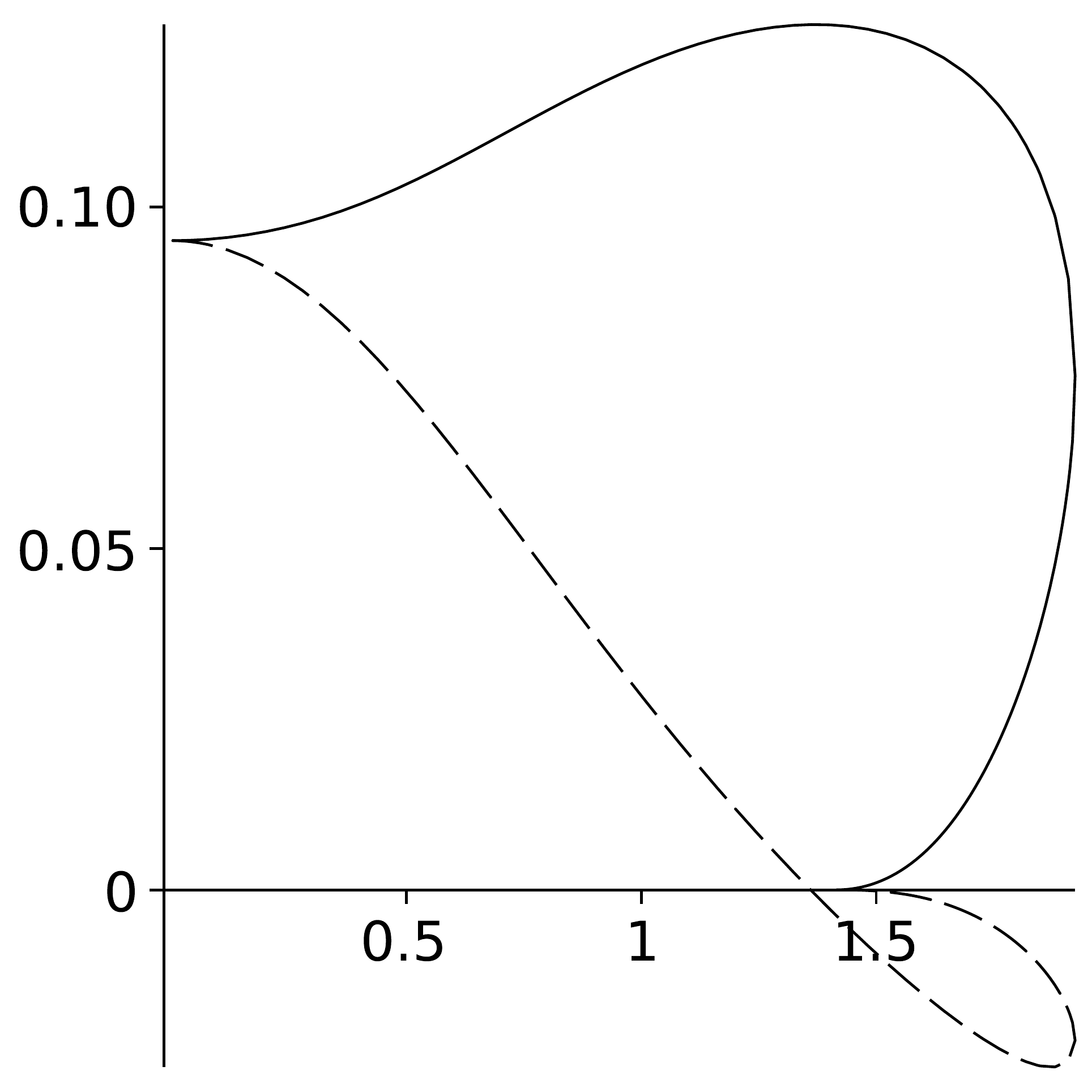}}
	\;
	\subfloat[]
	{\includegraphics[width=0.23\textwidth]{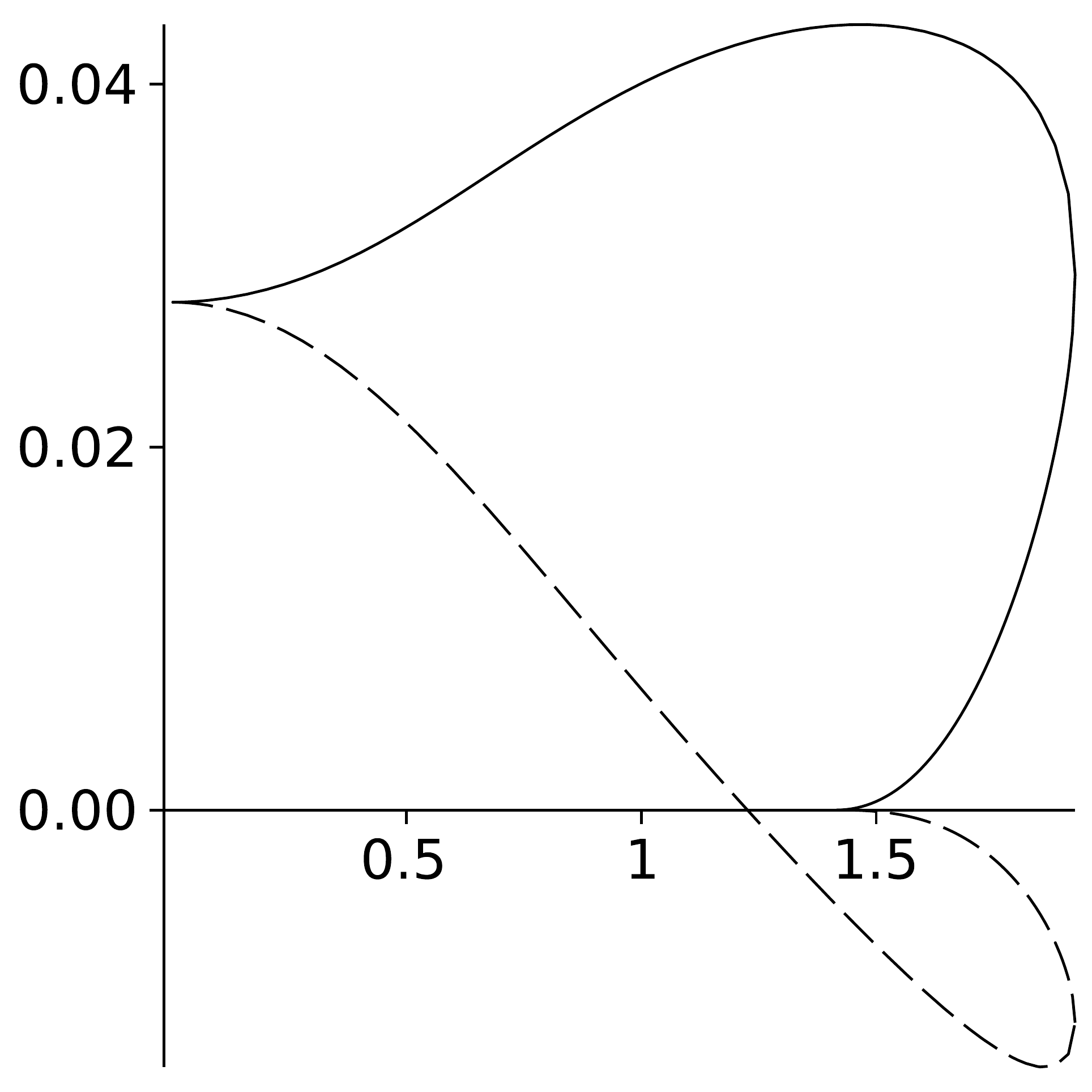}}
	\;
	\subfloat[]
	{\includegraphics[width=0.23\textwidth]{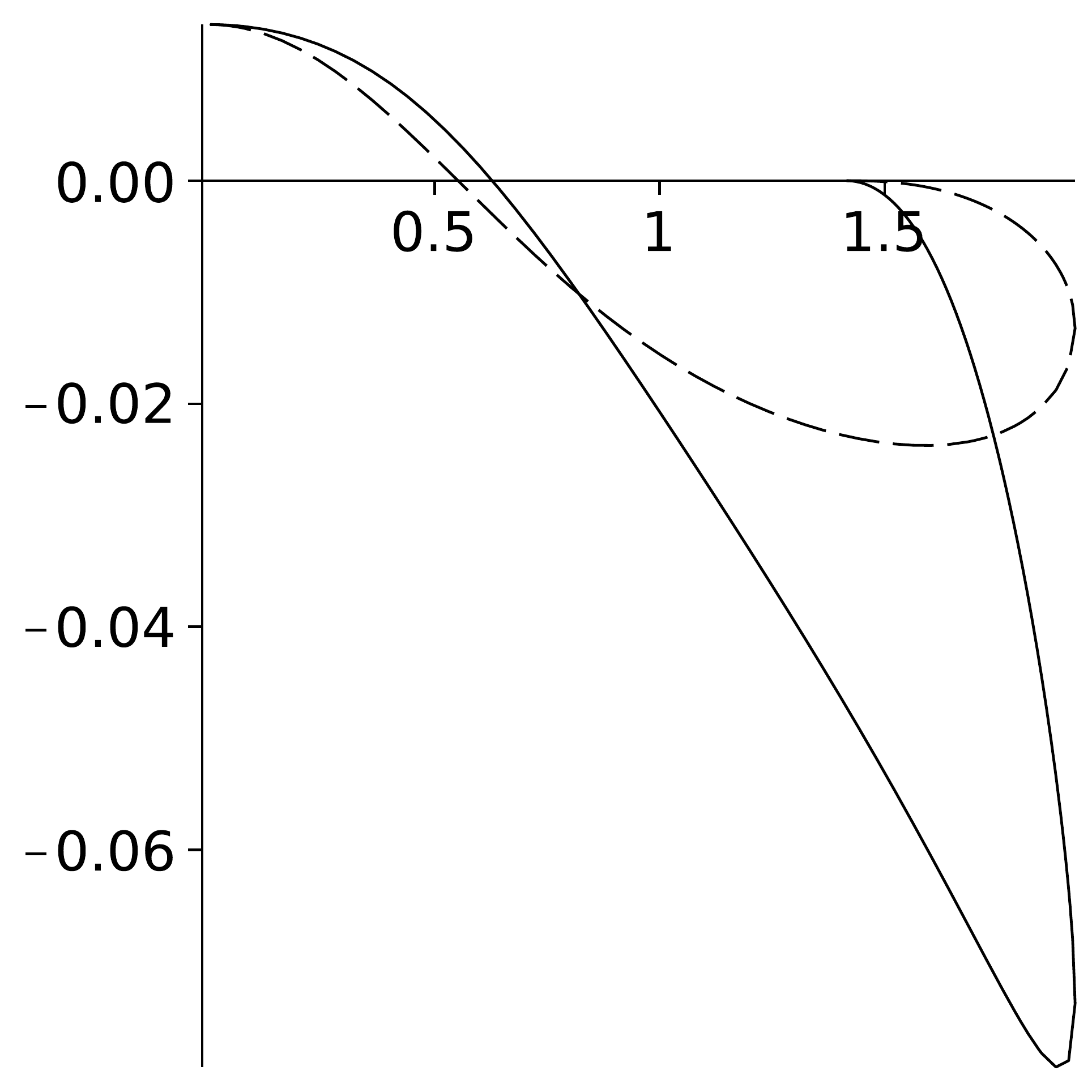}}
	\;
	\subfloat[]
	{\includegraphics[width=0.23\textwidth]{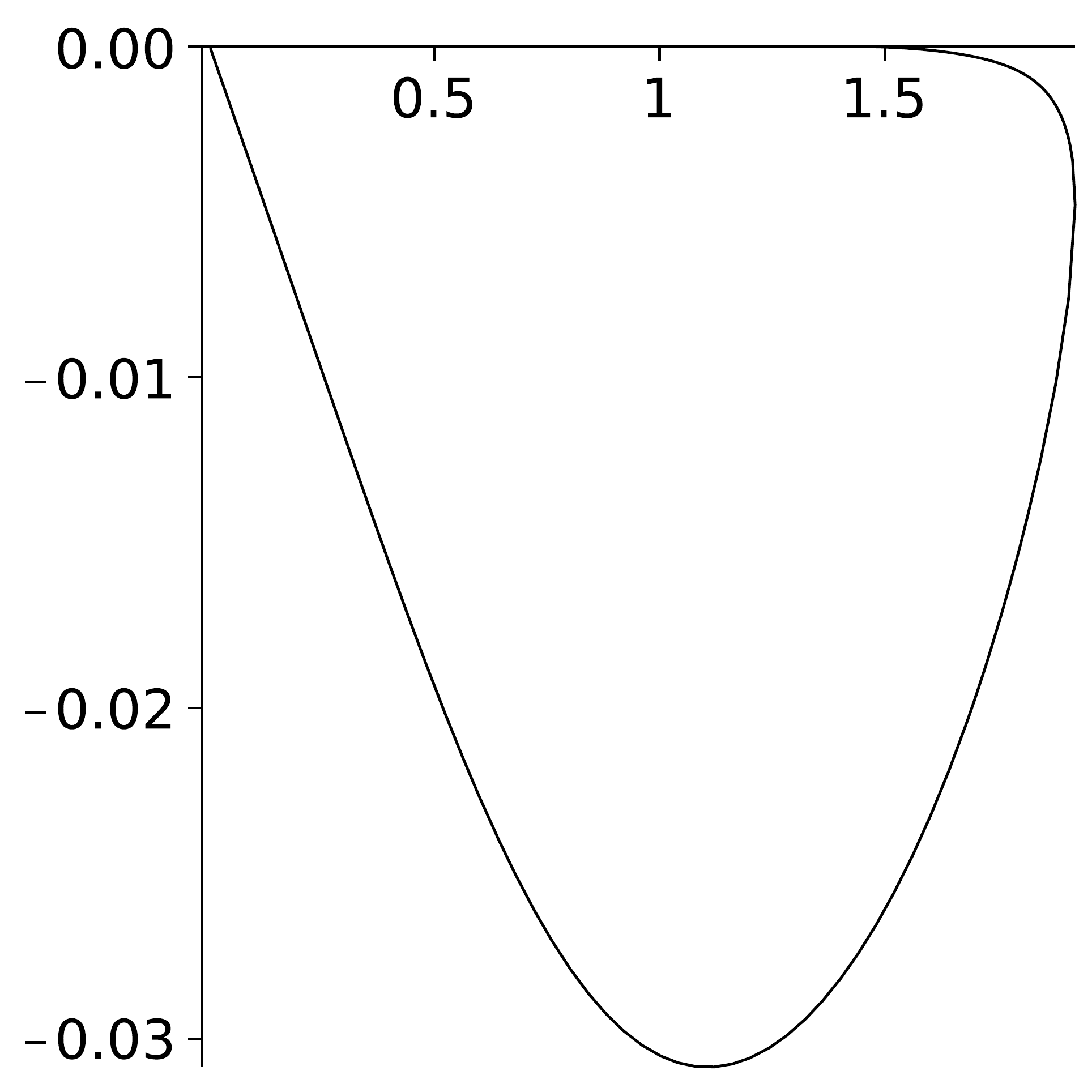}}
	\\
	\subfloat[]
	{\includegraphics[width=0.23\textwidth]{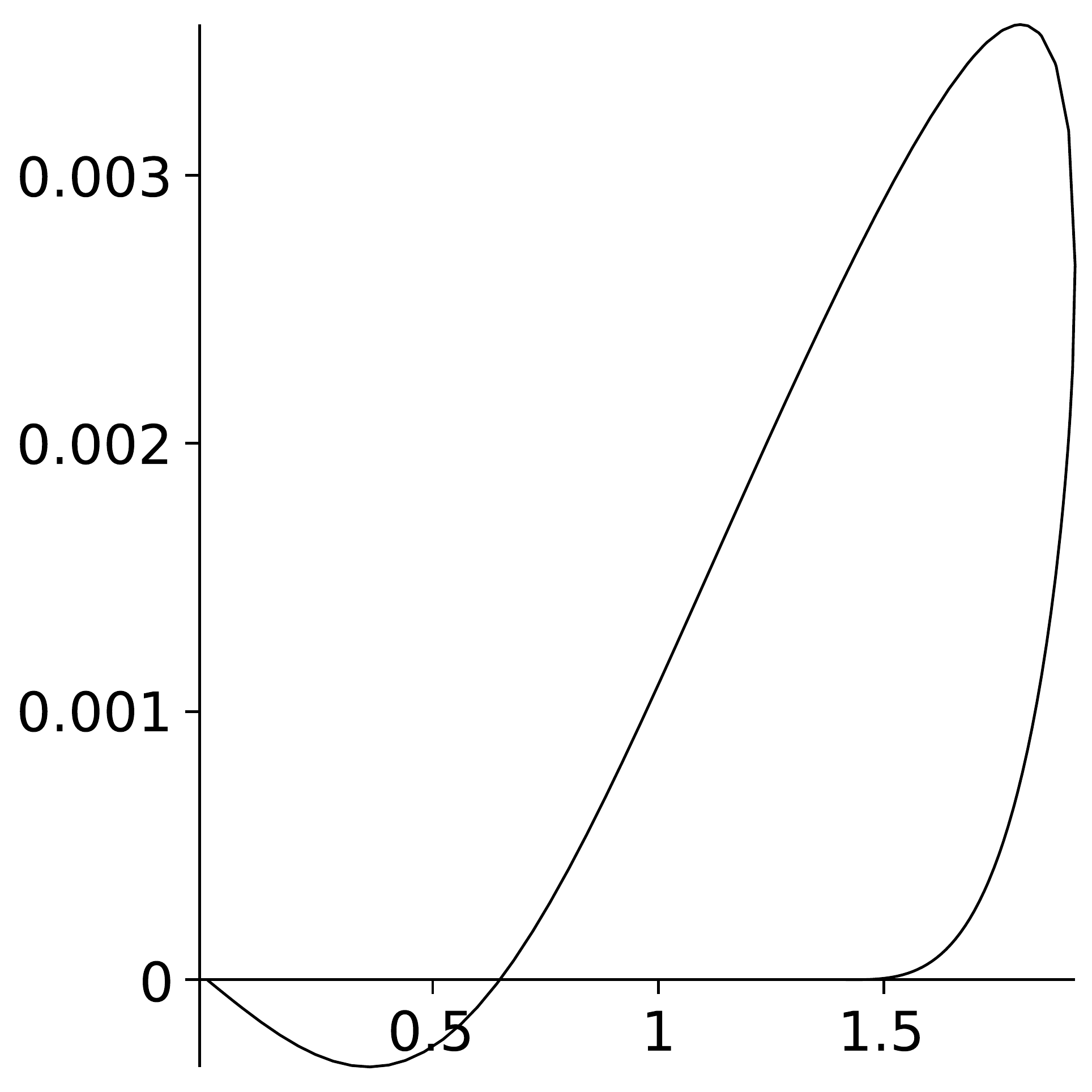}}
	\;
	\subfloat[]
	{\includegraphics[width=0.23\textwidth]{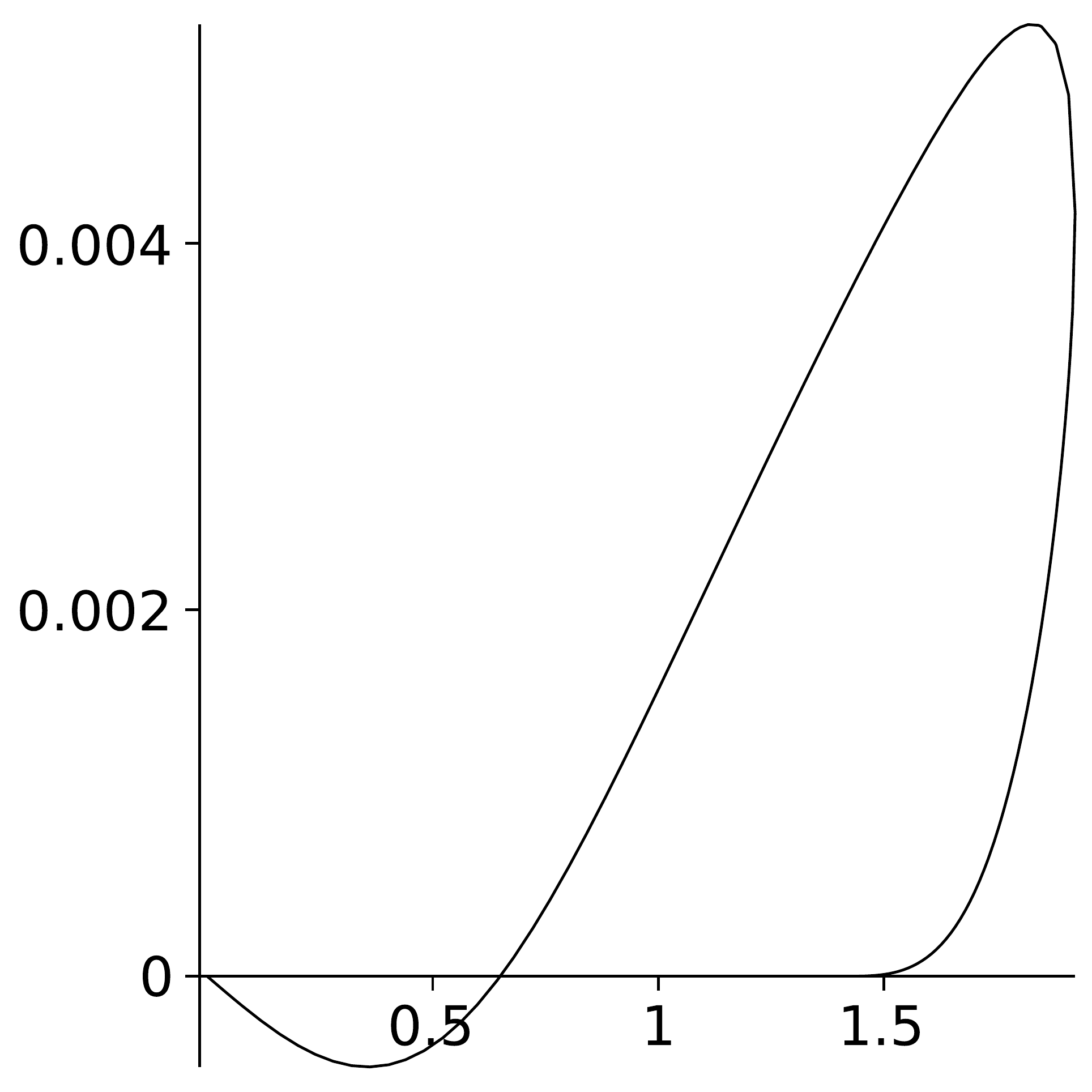}}
	\;
	\subfloat[]
	{\includegraphics[width=0.23\textwidth]{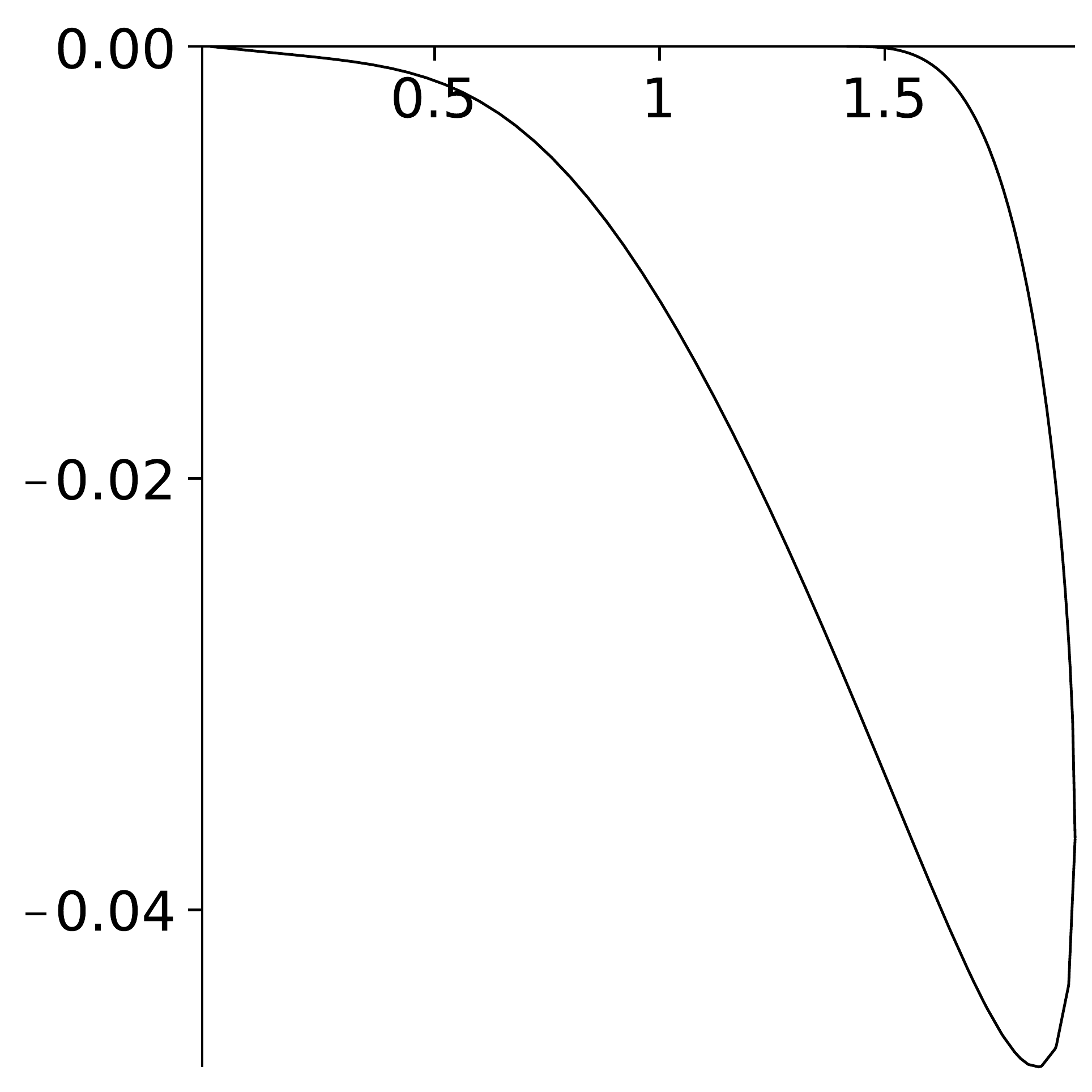}}
	\;
	\subfloat[]
	{\includegraphics[width=0.23\textwidth]{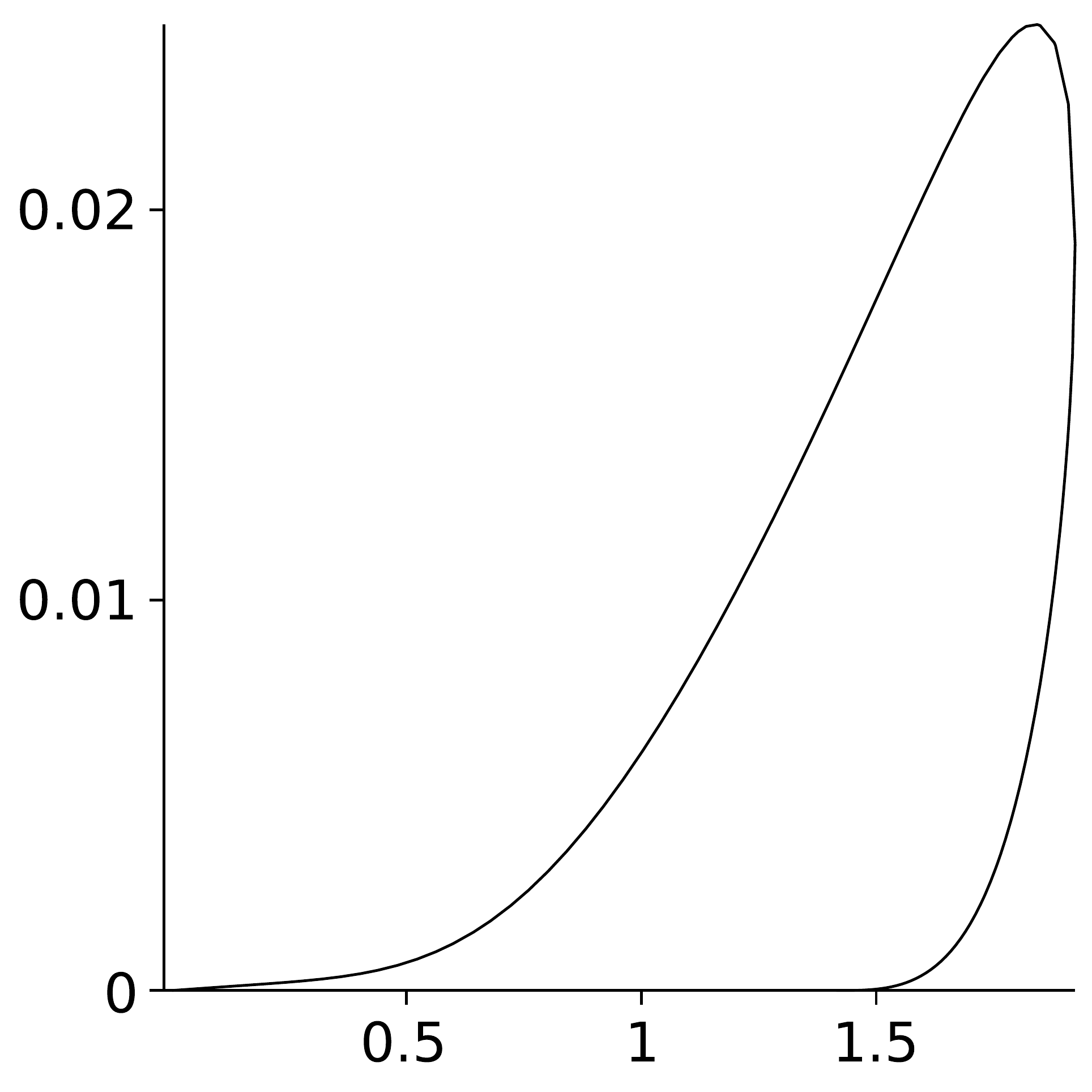}}
\caption{The other fields at level (3,9), as a function of initial
D-brane  separation, $d$.  (a) $U_s$ (solid line) and $\upsilon$ (dashed line).
(b) $V_{s}$ (solid line) and $\nu$ (dashed line).
(c) $\tilde{W}_{s}$ (solid line) and $\tilde{\omega}$ (dashed line).
(d) $\tilde{\phi}$.
(e) $R_{a}$.
(f) $S_{a}$.
(g) $Y_{a}$.
(h) $Z_{a}$.
}
\label{f:level3cont}
\end{figure}

At level 3, in addition to $t$ and $x$, we have the following 
twist-even, exchange-even fields:
\be\label{eq.2brane.exmatrices}\begin{gathered}
	u_{ij}=\begin{pmatrix}U_{s}&\upsilon\\ \upsilon&U_{s}\end{pmatrix},\quad
        v_{ij}=\begin{pmatrix}V_{s}&\nu\\ \nu&V_{s}\end{pmatrix},\quad
	\tilde w_{ij}=\begin{pmatrix}\tilde W_{s}&\tilde \omega\\ \tilde \omega&\tilde W_{s}\end{pmatrix},\quad
	\tilde f_{ij}=\begin{pmatrix}0&\tilde \phi\\ -\tilde \phi&0\end{pmatrix}, \\ 
	r_{ij}=\begin{pmatrix}-R_{a}&0\\ 0&R_{a}\end{pmatrix},\quad
	s_{ij}=\begin{pmatrix}-S_{a}&0\\ 0&S_{a}\end{pmatrix},\quad  
	y_{ij}=\begin{pmatrix}-Y_{a}&0\\ 0&Y_{a}\end{pmatrix},\quad
	z_{ij}=\begin{pmatrix}-Z_{a}&0\\ 0&Z_{a}\end{pmatrix}.
\end{gathered}\ee
We are using $\tilde w$ and $\tilde f$ instead of $w$ 
and $f$ (see Table \ref{string-field})
because, as explained in Section \ref{SF}, 
$w$ and $f$ are singular at $d=\sqrt 2$.
Figures \ref{f:level3} and \ref{f:level3cont}
show these fields, as well as the energy of the solution, 
as functions of the initial D-brane separation, $d$.

\begin{figure}
\center{\includegraphics[width=0.4\textwidth]{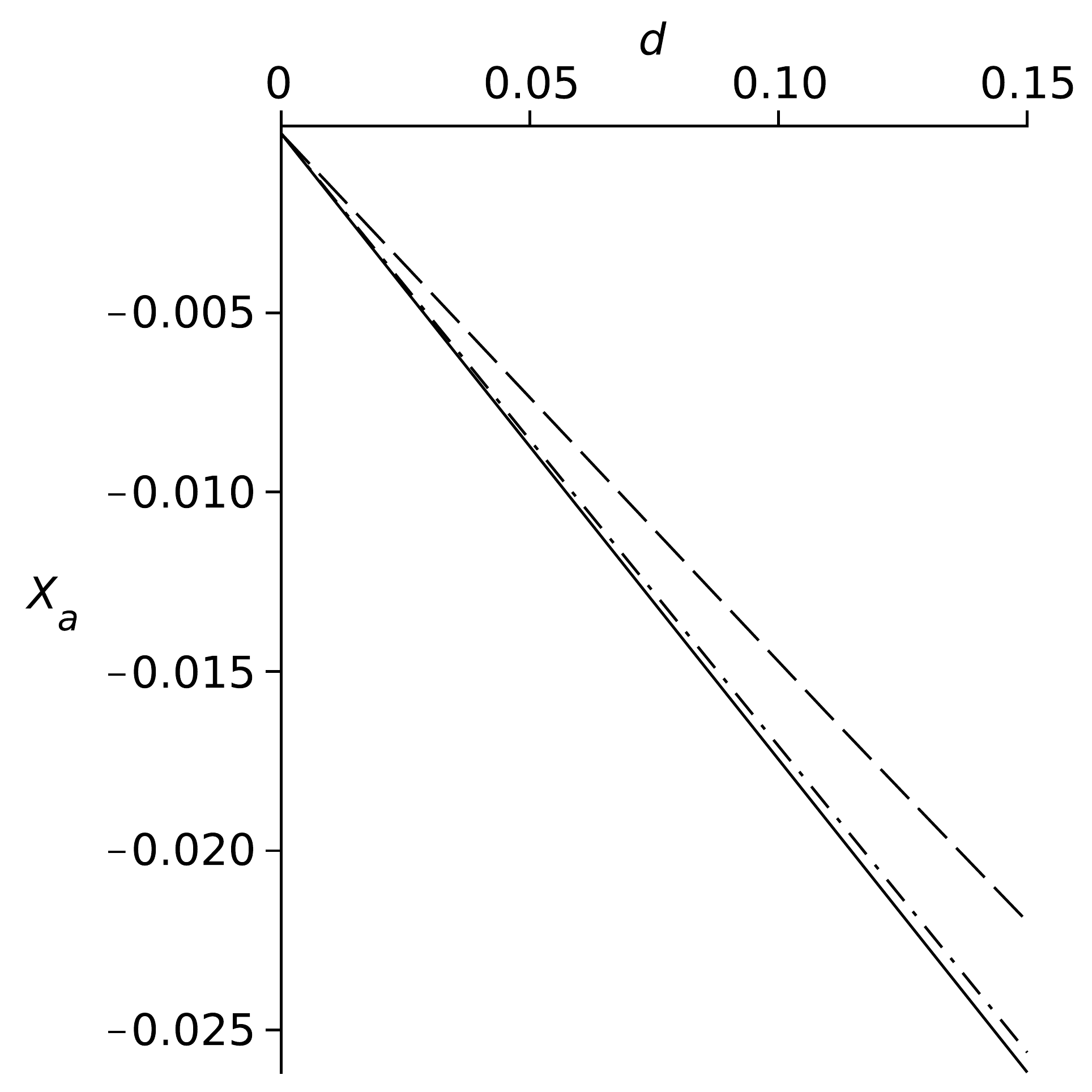}}
\caption{The field $X_a$ as a function
of the rescaled D-brane separation $d$ for small separations at different
truncation levels.  Level (3,9) is solid line and lower levels (2 and 1) have
shallower slopes.}
\label{f:slopes}
\end{figure}

\subsection{Discussion}
\label{decay:discussion}

We would like to interpret the solution we have found as representing a combination of 
a marginal deformation bringing the two D-branes to the same position and 
a tachyon condensation diagonal in some new basis.
This new basis is an SU(2) rotation of the original basis, possible
because SU(2) symmetry has been restored by the marginal deformation.
One should ask then why we only see one such solution 
(especially at level (1,3), where we have not imposed exchange symmetry
and where the full SU(2) family should be visible). The answer
is that level truncation does not allow for a full restoration of the SU(2) symmetry,
and only solutions at isolated points survive.  To see that level 
truncation affects the restoration of SU(2) consider two 
separated D-branes in the full theory.  There exists a solution in this 
theory corresponding to bringing the two D-branes together.  Expanding
the potential around this solution and performing a field 
redefinition\footnote{For  infinitesimal marginal deformations, 
these field redefinitions were computed in \cite{Sen:1993mh}.},
we should get back the SU(2) symmetric potential 
for fields living on two coincident
D-branes.  The field redefinition mixes fields at different levels and
is not the same for diagonal and off-diagonal elements,
because $X_a$ is nonzero while $\chi$ is not.  Fully
restoring SU(2) symmetry is therefore not possible in the truncated
theory. 

Several small comments on Figure \ref{f:level3} are in order.  There are 
two branches visible, one starting at zero separation, $d=0$, and the second one
starting at $d=\sqrt 2$ (the second branch starts at this point independent of
the level of truncation).  The two branches merge and end at $d \approx 1.92$
(beyond this point, the solutions become complex).  We attribute the
existence of the second branch to the fact that at $d=\sqrt 2$, the 
off-diagonal element of the tachyon field is massless.
The shape of the $X_a(d)$ curve (Figure \ref{f:level3d}) near the
point where the two branches meet cannot be determined at this
level of truncation: it might be a cusp or possibly even a loop.
In Figure \ref{f:level3c} we see that $T_s$ is no longer approximately equal to $\tau$.
This is due to the the marginal deformation component of the solution 
which has $T_s \neq \tau = 0$.
Figure \ref{f:level3b} shows the energy of the solution at different truncation
levels.  Surprisingly, the energy is somewhat less flat at levels 2 and 3 than it is at level 1.
The decrease in flatness when going from 
level 1 to level 2 might be related to the observation in \cite{Sen:2000hx}
that the leading quadratic term in the vacuum branch of the effective potential
for $a_s$, analogous to our field $X_{a}$, is larger at level 2 than at levels 1 or 3.
Because the marginal direction is lifted by level
truncation, we expect that the curves would become flatter again if the
truncation level were increased, despite the increased curvature
when we go from level 1 to level 2.

It is tempting to interpret the curves in Figure \ref{f:level3d} and in Figure
\ref{f:slopes} as corresponding
to the relationship between the vev of the marginal parameter in SFT, $X_a$, and 
the physical displacement of the D-branes, $d$. Unfortunately, this would not be correct,
even at linear level at small $d$, as seen in Figure \ref{f:slopes}.  
To see why, consider the  general form of the 
SFT (untruncated) potential with D-brane separation $d$, $f_d(X_a, \varphi^I, \xi^I)$, where
we have split the fields in this potential into three groups: the massless mode
$X_a$, all the other diagonal fields $ \varphi^I$ and the off-diagonal
fields $\xi^I$.  For $X_a$ to be massless, the potential must contain 
no $X_a^2$ term and no terms of the form $X_a \varphi_I$ or $X_a \xi_I$.
Further, no term can be linear or cubic in the fields $\xi^I$.  
On the other hand, the potential must contains terms of the form
$X_a^2 \phi^I$ whose coefficients do not depend on $d$.
Let $X_a = \bar X_a \neq 0$, $\varphi^I = \bar \varphi^I$, $\xi^I=0$
be a solution representing D-brane translation such that the two 
D-branes are coincident.   For small initial D-brane separations, 
$\bar X_a$ is simply proportional to $d$, but for larger separations their relationship
is more complicated.  Expanding around this solution, 
$X_a = \bar X_a + X'_a$, $\varphi^I = \bar \varphi^I + \varphi'^I$, 
$\xi^I=\xi'^I$ we obtain a potential for the new fields $X'_a$, $\varphi'^I$ and
$\xi'^I$.  With an appropriate field redefinition,$(X'_a, \varphi'^I, \xi'^I) \rightarrow
(\tilde X_a, \tilde \varphi^I, \tilde \xi^I)$, this potential
is equal to the potential at zero separation, $f_0$:
\be
f_d(\bar X_a + X'_a, \bar \varphi^I +  \varphi'^I, \xi'^I) = 
f_0(\tilde X_a, \tilde \varphi^I, \tilde \xi^I)
\ee
and the SU(2) symmetry is apparent.  We will work to leading order,
where the field redefinition is linear.
Thus, a particular linear combination of $X'_a$ and  $\varphi'^I$, 
$\tilde X_a \approx c_{XX} X'_a + \sum_I c_{XI} \varphi'^I $, appears massless in $f'$
(meaning that when written in terms of the redefined fields,
$f'$ has no term quadratic in $\tilde X_a$).   
Considering explicitly the expansion of the potential
$f_d(\bar X_a + X'_a, \bar \varphi^I +  \varphi'^I, \xi'^I)$ we see that it must
contain terms of the form $\bar X_a X'_a \varphi'_j$, with some $d$-independent
coefficients.  At small separations, $\bar X_a$
is proportional to $d$ and so this cross-term between $X'_a$ and  $\varphi'_j$ has a 
coefficient proportional to $d$.  This leads to mixing between $X'_a$ and  $\varphi'_j$
in the new massless eigenfield, $\tilde X_a$.  Explicitly,
as the separation between D-branes goes to zero, $d \rightarrow 0$,
$c_{XX} \rightarrow 1$ and $c_{XI} \sim d \rightarrow 0$.
Similarly, for $\tilde \varphi^I \approx c_{IX} X'_a + \sum_J c_{IJ} \varphi'^J $,
we have that $c_{II} \rightarrow 1$ while $ c_{IX}, c_{IJ} \rightarrow 0$ 
(for $I \neq J$).

In the new `tilde' variables, there is a SU(2) family of 
solutions representing the decay of a single D-brane.  These solution have $\tilde X_a = 0$,
while $\tilde \varphi^I$ are nonzero and do not depend on the initial D-brane separation.
Thus, $X'_a$ is non-zero, unless there is some cancellation, which we have
no reason to expect.  More explicitly, at small $d$, we have that $\varphi'^J \approx \tilde 
\varphi^J$, and $0 = \tilde X_a \approx X'_a + \sum_I c_{XI} \varphi'^I$ so that
$X'_a \approx  - \sum_I  c_{XI} \tilde \varphi^I$.  Since we already argued that
$c_{XI}$ decrease linearly with $d \rightarrow 0$, this implies that
$X'_a$ is also linear in $d$.  Therefore, in the combined translation-and-decay
solution, the vev of the massless mode is $X_a = \bar X_a + X'_a$, 
is not the same as it would be with translation alone.  The
correction, $X'_a$, is of the same order in $d$ as $\bar X_a$ itself, so
the vertical axis in Figure \ref{f:level3d} does not represent the vev of the marginal parameter
responsible for a translation, even at small $d$.  To be able to understand in
detail the relationship between the vev of the massless SFT field
and the vev in the CFT from our solutions, we would need to understand the field redefinition
between $\varphi'$ and $\tilde \varphi$.  We leave this problem for future work,
but point out that solving it requires only a better understanding of the marginal
deformation in SFT and not of any connections between SFT and the CFT.

Still, because it was computed by controlling the D-brane displacement itself
as the adjustable parameter, and not a parameter in the string field, 
Figure \ref{f:level3d} contains a very 
interesting piece of information: There is a finite maximum D-brane 
separation $d \approx 1.9$ beyond which the solutions
do not exist.  This corresponds to a physical separation between the D-branes equal to
$\pi \sqrt {2\alpha'} d \approx 8.5 \sqrt {\alpha'}$.
The implication is that open string field theory in this particular coordinate system
is unable to describe the displacement of a D-brane beyond half this distance, and therefore
fails to describe the full CFT moduli space.

This answers the question raised in \cite{Sen:2000hx}.
In that paper, a marginal deformation is studied by assuming a predetermined
value for a certain marginal parameter
in the string field ($a_s$ in \cite{Sen:2000hx}, T-dual to our $X_a$),
solving the equations of motion of all \emph{other} fields
and thus computing the effective potential for $a_s$.  Level truncation lifts
this potential and what is obtained is not truly a solution to the complete
string field theory equations of motion, as the equation of motion for $a_s$ is
not satisfied.  It is found there that even the equations
of motion for the remaining fields cannot be solved at all once $a_s$ is greater than some critical
value $\bar a_s$, and that at $a_s = \bar a_s$ the `solution'
in merges with another branch.  The value of $\bar a_s$ computed in \cite{Sen:2000hx} 
is about 0.331 (at level (4,8)). In contrast, we find \emph{actual} solutions to the truncated
equations of motion, but our solution is a combination of the marginal deformation
and a decay of one of the two D-branes.  We also find that there are 
no solutions beyond a certain point; the largest $|X_a|$ attained for our solutions is
$0.1557$ (level 1), $0.2431$ (level 2) and $0.2579$ (level 3).  As we have already
discussed, this is not the actual marginal parameter, so we cannot compare our
values with those of \cite{Sen:2000hx}, though we note they are of the same order of
magnitude.  Qualitatively we do see the same phenomenon: the marginal deformation has a finite
range.  However, since we have the physical distance through which the D-branes have 
been displaced, we can also say that this finite range of marginal
deformation parameter corresponds to a finite range of the CFT vev,
which the authors of \cite{Sen:2000hx} were unable to do.  

It is interesting to ask whether the finite range of defomation can be an artifact of
either trucation or breakdown of Siegel gauge.  In Figure
\ref{f:level3b} we see that there is some indication of covergence with increased level,
including convergence of the range over which the deformation exists.  It would be 
quite interesting to explore the trucation at higher levels.  In particular,
it might be that increasing the level from even to odd has a smaller effect than
increasing the level from odd to even.  If that is the case, a computation at level
4 could be quite telling.  The question about gauge
validity is hard to settle given our data.  While the value of tachyon
field is well within the region where Siegel gauge holds \cite{Moeller:2000xv,Ellwood:2001ne} for the entire
solution including the brach point, (Figure \ref{f:level3}(c)), it is possible
that the breakdown of Siegel gauge occurs at a smaller tachyon field in our set up.
It would be interesting to explore this possibility, by computing the effective
potential for the tachyon at different brane separations and repeating the analysis of
\cite{Ellwood:2001ne} for multiple separated D-branes.

An interesting interpretation of the results in \cite{Sen:2000hx} was given in
\cite{Zwiebach:2000dk}: there it was proposed that there is another branch
of solutions to the SFT equations of motion, so that, at the same value of the marginal
deformation parameter there can be two solutions, differing in the higher level fields,
representing two different vevs in the CFT.  This would allow OSFT to cover the full
CFT moduli space.  We find no indication in our computation of the existence of such a
branch.

\section{Restoration of SU(2) symmetry}
\label{su2}

In the previous section, we used the presence of an off-diagonal tachyon 
condensate as a signal that the two D-branes have been brought together 
and SU(2) symmetry has been restored.  In this section,
we discuss an attempt to find solutions corresponding to the SU(2)
symmetric point directly, by examining the spectrum of the theory around an
approximate solution corresponding to a purely marginal deformation.

Starting with two spatially separated D-branes, there should
exist a solution in the untruncated SFT which corresponds to bringing these two D-branes
together, restoring  SU(2) symmetry.  The SFT action for
small fluctuations around such a string field will have an explicit
SU(2) symmetry, reflected, for example, in the degeneracies of the mass
spectrum for these small fluctuations.  Unfortunately,
due to mixing of fields at different levels, in a level-truncated model
the symmetry is restored only approximately.  

\begin{figure}
	\centering
	\subfloat[]
	{\includegraphics[width=0.48\textwidth]{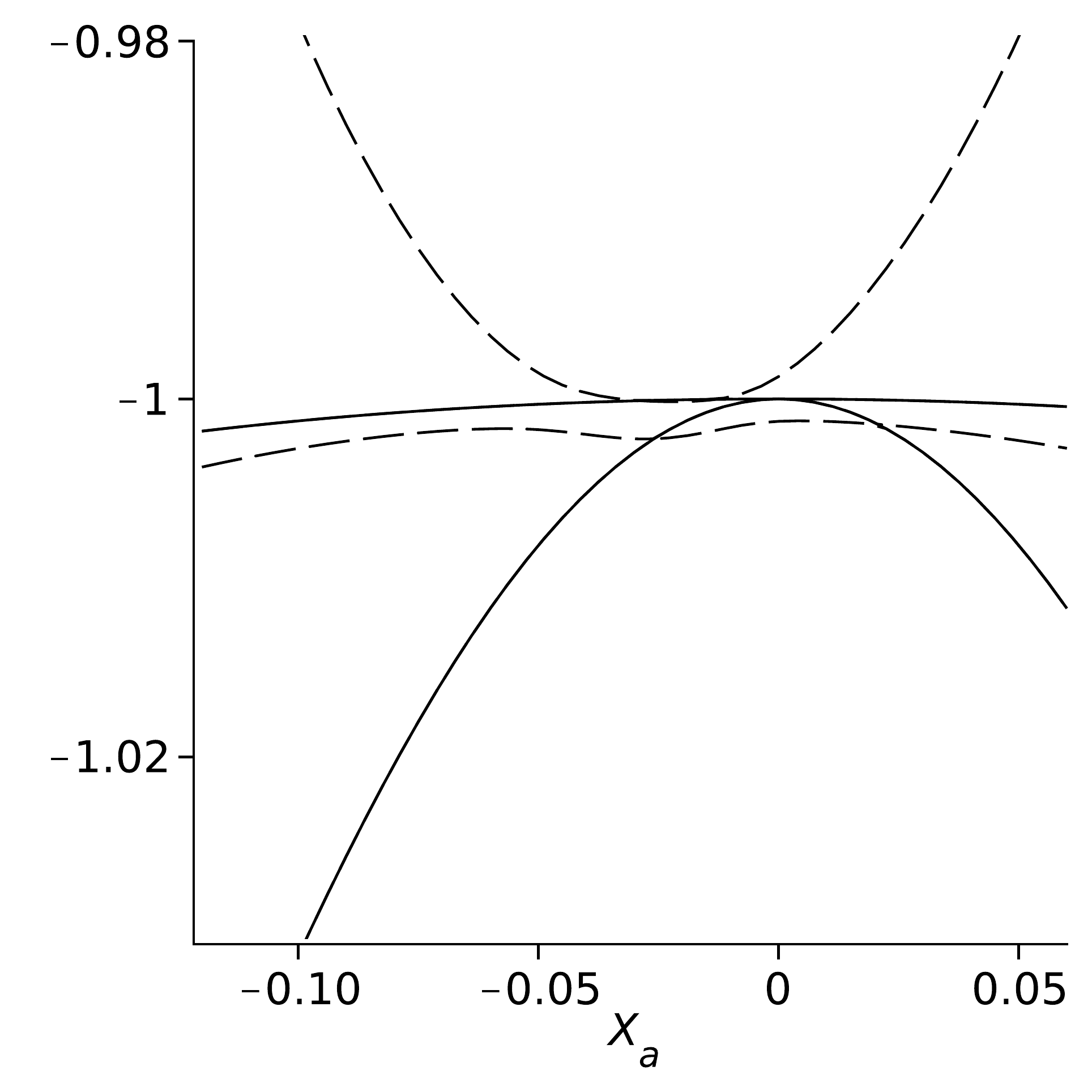} \label{fig.lmd-tu-eigen005}}
	\subfloat[]
	{\includegraphics[width=0.48\textwidth]{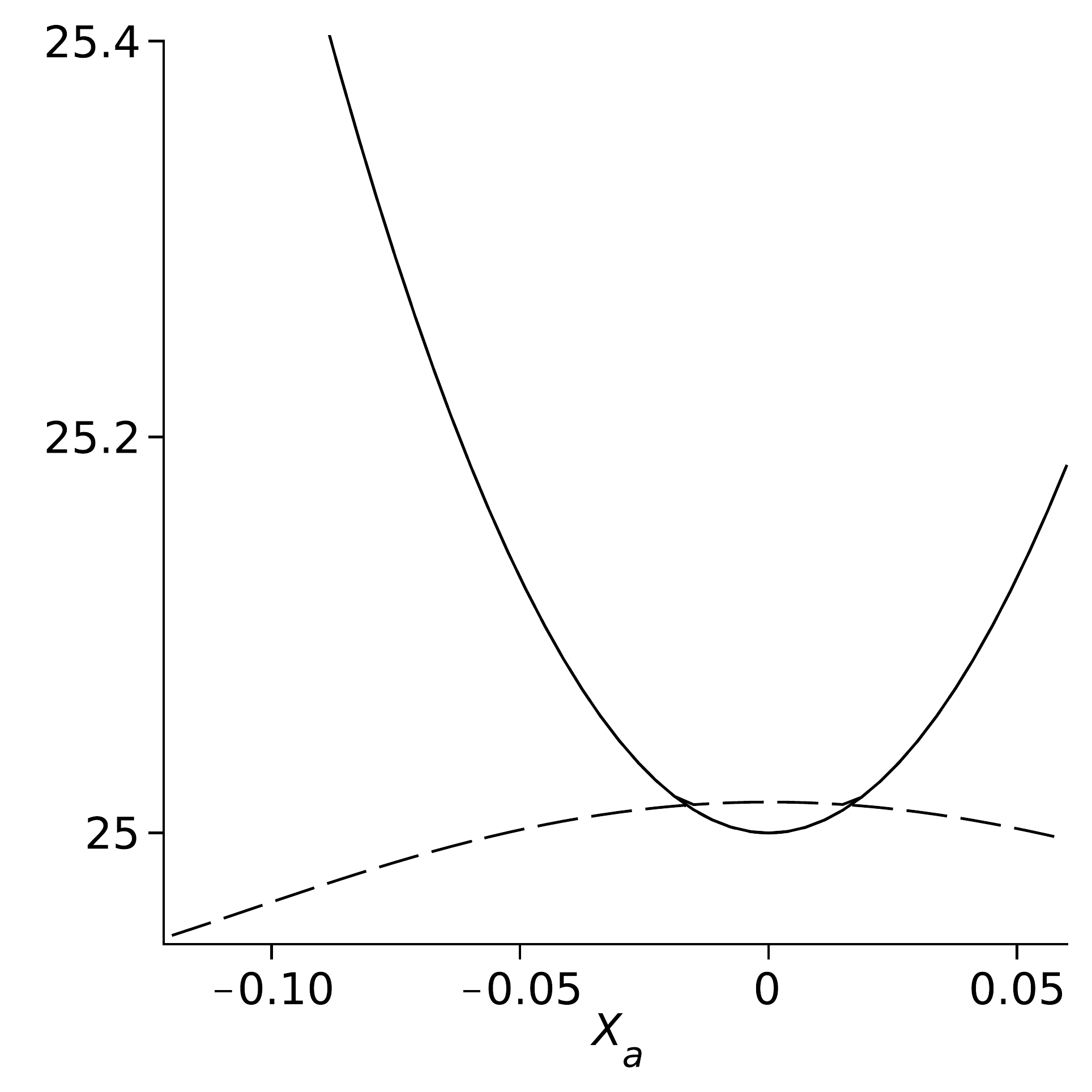} \label{fig.lmd-s-eigen005}} \\
	\subfloat[]
	{\includegraphics[width=0.48\textwidth]{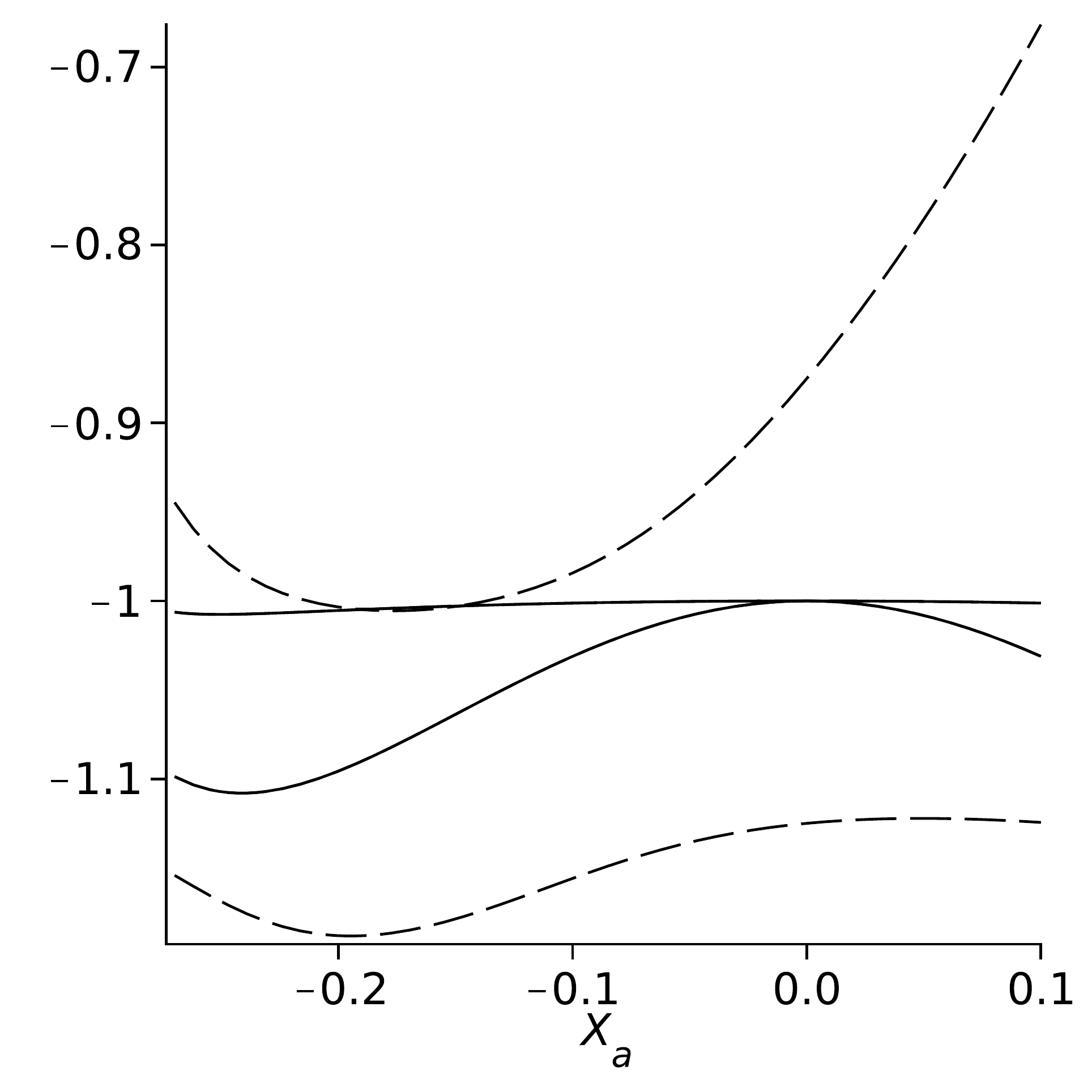} \label{fig.lmd-tu-eigen05}}
	\subfloat[]
	{\includegraphics[width=0.48\textwidth]{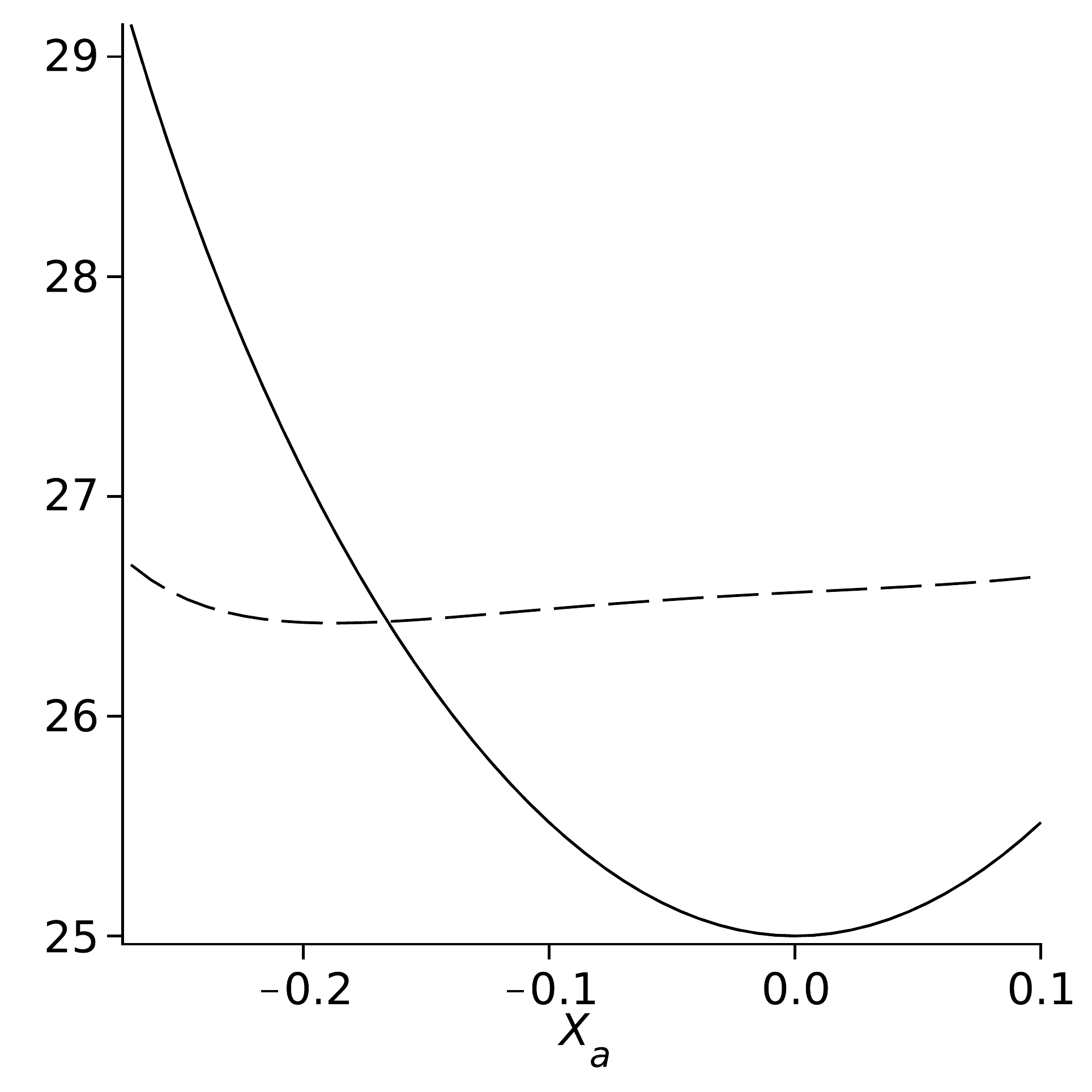} \label{fig.lmd-s-eigen05}}
	\caption{Eigenvalues of the matrix of second derivatives of the potential
near an approximate solution for a fixed $X_a$, as a function of $X_a$.  The initial
D-brane separation corresponds to $d=0.05$ in (a) and (b) and to $d=0.5$ in (c) and (d). 
(a),(c) Eigenvalues corresponding to the fields
$t_{ij}$ and $u_{ij}$. (b),(d) Eigenvalues corresponding to the fields $s_{ij}$. Dashed lines indicate a 
non-degenerate eigenvalue while solid lines correspond to doubly-degenerate eigenvalues.}
\label{fig.eigen}
\end{figure}

Our strategy is as follows: starting with a theory with a given D-brane
separation $d$, we construct a one parameter set of approximate solutions corresponding
to different marginal deformations bringing the two D-branes closer together by
a varying amount.
These approximate solutions are constructed using the approach in \cite{Sen:2000hx},
{\it i.e.} by picking a marginal deformation parameter $X_a$ and solving the 
equations of motion for all the other fields in the potential.  Once we have the 
approximate solution, we find the matrix of second derivatives of the potential
w.r.t. all the fields and diagonalize it.  If we were to perform this computation
in the untruncated theory, we would expect to find that at a particular value
of $X_a$, $\bar X_a(d)$, the spectrum would
develop degeneracies at the point where the SU(2) symmetry is restored.
In addition, near the degenerate point, the spectrum as a function of
the $X_a$ should be reflection-symmetric about $X_a = \bar X_a(d)$---bringing 
the D-branes nearly together should produce the 
same spectrum as `overshooting' a bit.  Identifying the degenerate point 
for different values of $d$ would produce the function $\bar X_a(d)$,
describing the relationship between the strength of the OSFT deformation
$\bar X_a$ and the CFT vev it produces, $d/2$.

In a level truncated theory, we would hope that this degeneracy is present at 
least approximately.  Note that while we focus on the matrix of second
derivatives, in level truncation, the first derivatives of the potential
do not all vanish, since the equation of motion for $X_a$ is not satisfied.
This effect decreases with increased truncation level.

Our results, for the eigenvalues corresponding to the masses of selected fields are shown in 
Figure \ref{fig.eigen}. The computation used twist-even fields only.  The approximate solution is 
also exchange-even, and we included all twist-even (both exchange-even and exchange-odd) fields
in the computation of the second derivative matrix.  At zero separation, the fields $t$ and $u$ 
(\ref{fig.lmd-tu-eigen005},\ref{fig.lmd-tu-eigen05}) have masses equal $-1$ while the fields in $s$ have mass $25$
(\ref{fig.lmd-s-eigen005},\ref{fig.lmd-s-eigen05}) (see Appendix).  
Unfortunately, the features we just discussed do not seem to be 
unambiguously visible at level (3,9).
Apparently, the cubic couplings to higher level 
fields with non-zero vev contribute nontrivially 
to the masses of the lower modes when the D-branes are translated.
It would be interesting to see whether this can be improved at higher levels.

\section*{Appendix: The potential at higher levels}
\addcontentsline{toc}{section}{Appendix: The potential at higher levels}

Let us denote our set of string fields, $t$, $x$, $u$, $\ldots$, with $\phi^{(m)}$.
Then the string field potential (\ref{eq.sep.basic-action}) can be written as
\bear \label{eq.appendix.pot-defn}
f&=&\pi^{2}\sum_{l,m}~\sum_{ij}~A_{lm}(d_{ij})\phi^{(l)}_{ij}\phi^{(m)}_{ji} \\ \nn
&-&2\pi^{2}\sum_{l,m,n}~\sum_{ijk} ~B_{lmn}(d_{ij},d_{jk},d_{ki})
\left(\frac{4}{3\sqrt{3}}\right)^{\frac{1}{2}(d_{ij}^{2}+d_{jk}^{2}+d_{ki}^{2})}\phi^{(l)}_{ij}\phi^{(m)}_{jk}\phi^{(n)}_{ki},
\eear
From equation (\ref{quadratic-coefficient}) we get that $A_{lm}=\beta_m \beta_l A_{ml}$.
The symmetry properties of the coefficients $B_{lmn}(d_{ij},d_{jk},d_{ki})$ were
discussed in section \ref{SF} (where these coefficients were denoted with
$g(d_{ij}, d_{jk}, d_{ki})$).  Notice that the parameters $d_{ab}$ appearing in
the coefficients $A_{lm}(d_{ij})$ and $B_{lmn}(d_{ij},d_{jk},d_{ki})$ are just
the eigenvalues of $\alpha_0^{25}$ in the lowest state of each $ab$ sector of our theory.
These can have a parallel  interpretation in other scenarios, such as
string theory on a single D-brane on a circle.  This allows us to
compare some of our coefficients to those computed for example in \cite{Moeller:2000jy}.

The coefficients $A_{lm}(d)$ for the quadratic part of the potential appear in 
Table \ref{table-quadratic}.
The string fields $o$ and $\tilde o$ are defined by $o_1 = o + \tilde o$ and $o_2 = 2 \tilde o - 2 o$.
The coefficients $B_{lmn}(d_{1},d_{2},d_{3})$ for the cubic part of the potential up to level (3,9) 
have the form of polynomials in ($d_1$, $d_2$, $d_3$) for example:
\bear
B_{xxw}(d_{1},d_{2},d_{3}) &=& 
-{\frac {1}{864}}\,\sqrt {3} ( 4\,d_{{1}}{d_{{2}}}^{3}-108+12\,{d
_{{3}}}^{3}d_{{2}}+12\,{d_{{3}}}^{3}d_{{1}}-37\,d_{{1}}d_{{2}}-24\,{d_
  {{3}}}^{2}d_{{2}}d_{{1}} \nn \\ &-& 8\,{d_{{1}}}^{2}{d_{{2}}}^{2}+4\,d_{{2}}{d_{{
1}}}^{3}-4\,d_{{3}}{d_{{2}}}^{3}+155\,{d_{{3}}}^{2}+16\,{d_{{1}}}^{2}+
16\,{d_{{2}}}^{2}+8\,d_{{3}}d_{{2}}{d_{{1}}}^{2}\nn\\&-&8\,{d_{{3}}}^{4}-75\,
d_{{3}}d_{{1}}-75\,d_{{3}}d_{{2}}-4\,d_{{3}}{d_{{1}}}^{3}+8\,d_{{3}}d_
{{1}}{d_{{2}}}^{2} ) ~.
\eear
A full set of these coefficients will appear elsewhere \cite{Longton:2012ei}.

\begin{table}[h]\begin{center}\begin{tabular}{@{}|c@{}c|l|}
\hline
$l$ & $m$ & $A_{lm}(d)$ \\ \hline

$t$ & $t$ & $1/2\,{d}^{2}-1
$ \\ 
$x$ & $x$ & $1/2\,{d}^{2}
$ \\ 
$h$ & $h$ & $-2
$ \\ 
$u$ & $u$ & $-1-1/2\,{d}^{2}
$ \\ 
$v$ & $v$ & ${\frac {25}{2}}+{\frac {25}{4}}\,{d}^{2}
$ \\ 
$w$ & $w$ & $1/4\, ( 4\,{d}^{2}+1 )  ( {d}^{2}+2 ) 
$ \\ 
$w$ & $f$ & $3/2\,d ( {d}^{2}+2 ) 
$ \\ 
$f$ & $w$ & $-3/2\,d ( {d}^{2}+2 ) 
$ \\ 
$f$ & $f$ & $- ( {d}^{2}+2 )  ( {d}^{2}+1 ) 
$ \\ 
$o$ & $o$ & $8+2\,{d}^{2}
$ \\ 
$\tilde{o}$ & $\tilde{o}$ & $-8-2\,{d}^{2}
$ \\ 
$p$ & $p$ & $-100-25\,{d}^{2}
$ \\ 
$q$ & $q$ & $-1/2\, ( 3\,{d}^{2}+2 )  ( 4+{d}^{2} ) 
$ \\ 
$q$ & $y$ & $-5/2\,d ( 4+{d}^{2} ) 
$ \\ 
$y$ & $q$ & $5/2\,d ( 4+{d}^{2} ) 
$ \\ 
$q$ & $z$ & $-6\,d ( 4+{d}^{2} ) 
$ \\ 
$z$ & $q$ & $6\,d ( 4+{d}^{2} ) 
$ \\ 
$r$ & $r$ & $-2-1/2\,{d}^{2}
$ \\ 
$s$ & $s$ & ${\frac {25}{4}}\,{d}^{2}+25
$ \\ 
$y$ & $y$ & $1/4\, ( 4+{d}^{2} )  ( 4\,{d}^{2}+9 ) 
$ \\ 
$y$ & $z$ & $3/2\, ( 3\,{d}^{2}+2 )  ( 4+{d}^{2} ) 
$ \\ 
$z$ & $y$ & $3/2\, ( 3\,{d}^{2}+2 )  ( 4+{d}^{2} ) 
$ \\ 
$z$ & $z$ & $3\, ( 4+{d}^{2} )  ( {d}^{2}+2 )  ( {d}^{2}
+1 ) 
$ \\ 
\hline
\end{tabular}\end{center}
\caption{The quadratic coefficients $A_{lm}(d)$  up to level 3.
Omitted coefficients are zero.}
\label{table-quadratic}
\end{table}

%%%%%%%%%%%%%%%%%%%%%%%%%%%%%%%%%%%%%%%%%%%%%%%%%%%%%%%%%%%%%%%%%%%%
%  END MATTER: BIBLIOGRAPHY, ACKNOWLEDGMENTS, ...                  %
%%%%%%%%%%%%%%%%%%%%%%%%%%%%%%%%%%%%%%%%%%%%%%%%%%%%%%%%%%%%%%%%%%%%

\section*{Acknowledgments}
This work was completed with support from the Natural Sciences and Engineering
Council of Canada.

\bibliographystyle{JHEP}
\bibliography{my.sft}

\end{document}